# Frustrated metastable-to-equilibrium grain boundary structural transition in NbMoTaW due to segregation and chemical complexity

Ian Geiger [1], Diran Apelian [1,2], Xiaoqing Pan [2,3], Penghui Cao [4], Jian Luo [5,6], Timothy J. Rupert [1,2,4,*]

[1] Material and Manufacturing Technology, University of California, Irvine, CA 92697, USA
[2] Department of Materials Science and Engineering, University of California, Irvine, CA 92697, USA
[3] Department of Physics and Astronomy, University of California, Irvine, CA 92697, USA
[4] Department of Mechanical and Aerospace Engineering, University of California, Irvine, CA 92697, USA
[5] Department of NanoEngineering, University of California San Diego, La Jolla, CA 92093, USA
[6] Program in Materials Science and Engineering, University of California San Diego, La Jolla, CA 92093, USA
* Corresponding Author: trupert@uci.edu

**Abstract**

Grain boundary structural transitions can lead to significant changes in the properties and performance of materials. In multi-principal element alloys, understanding these transitions becomes complex due to phenomena such as local chemical ordering and multi-component segregation. Using atomistic simulations, we explore a metastable-to-equilibrium grain boundary structural transition in NbMoTaW. The transition, characterized by structural disordering and reduced free volume, shows high sensitivity to its local chemical environment. Most notably, the transition temperature range of the alloy is more than twice that of a pure metal. Differences in composition between coexisting metastable and equilibrium structures highlight the change in local site availability due to structural relaxation. Further examination of grain boundaries with fixed chemical states at varying temperatures reveals that the amount of segregation significantly influences the onset temperature yet has minimal effect on the transition width. These insights underscore the profound effect of chemical complexity and ordering on grain boundary transitions in complex concentrated alloys, marking a meaningful advancement in our understanding of grain boundary behavior at the atomic level.

## 1. Introduction

Grain boundaries are essential in determining the overall behavior of polycrystalline materials [1,2], with their local structure and chemistry significantly influencing a range of mechanical [3–5], thermal [6,7], and electrical [8,9] properties. Recent advances in interfacial science have revealed the phase-like behavior of grain boundaries, indicating that a multitude of microscopic interfacial states are available for a given macroscopic description [10]. To access these unique states, transitions between stable and/or metastable states can be enabled and studied by varying thermodynamic parameters such as temperature ($T$), pressure ($P$), or chemical potential ($\mu_i$). For example, discontinuities in the interfacial solute excess can trigger grain boundary structural transitions due to the influence of bonding state on interfacial structure [3,11–13]. Despite the common focus on equilibrium interfacial states, the wide energetic dispersion and degeneracy of metastable grain boundary structures, as demonstrated by Han et al. [14] using MD, suggests that transitions between non-equilibrium states must also occur regularly as a result of processing conditions or dynamic processes. For example, annealing-induced relaxation of metastable grain boundaries in nanocrystalline metals is a common approach for pushing interfacial structure closer to equilibrium, resulting in significant changes to radiation tolerance [15], hardness [16], strength [17], and ductility [18]. The role of solute enrichment further complicates non-equilibrium grain boundary structures and their stability against transitions [10]. Grain boundaries with higher disorder (e.g., higher excess volume) typically exhibit greater solute saturation [19], leading to an increased activation energy barrier for transitioning to a more ordered structure with less enrichment [20]. Chemical and mechanical trapping of metastable interfaces are thus fundamental features governing interfacial evolution, emphasizing the need for understanding of metastable-to-equilibrium grain boundary structural transitions.

The investigation of grain boundary transitions has predominantly focused on unary and binary systems, but there is a growing need to understand such interfacial phenomena in more chemically complex materials. Multi-principal element alloys (MPEAs), for example, show great promise as next-generation engineering alloys [21,22], demonstrating properties such as excellent high temperature strength [23–25], outstanding wear resistance [26,27], and exceptional fatigue and fracture resistance [28,29]. Initially referred to as "high-entropy alloys" due to the anticipated role of mixing entropy in single-phase stability [30], the distinctive properties of MPEAs are now attributed to a rich landscape of local chemical domains, encompassing phenomena such as secondary-phase formation [31,32], nanoprecipitation [33], and chemical ordering [34,35]. These factors have broadened alloy design parameters beyond initial considerations. For example, in body-centered cubic (BCC) TiZrHfNb, the affinity of oxygen interstitials for regions with Ti-Zr chemical short-range ordering (CSRO) induces significant local lattice distortion, resulting in both interstitial strengthening and enhanced ductility through a modified dislocation propagation mechanism [36]. Tuning the concentrations of Nb and O in Ti-30Zr-xNb-yO alloys further amplifies the formation of these oxygen complexes [37], highlighting the impact and adjustability of local chemical domains in MPEAs. These findings motivate further exploration into the role of chemical complexity on defect behavior.

Besides the crystalline regions inside the grains, the immense configurational phase space in MPEAs presents new challenges in predicting the complex chemical and structural relationships at grain boundaries. Unlike dilute alloys, the solute/solvent relationship is less clear in MPEAs, complicating predictions of interface stability due to multiple segregating elements competing for similar sites, interacting with one another, or altering the local structure and therefore site availability [38–41]. Phenomenological thermodynamic models have been used to predict some trends in grain boundary segregation and disordering in MPEAs [41,42], as well as related properties (e.g., sintering from pre-melting like grain boundary transitions [43]). Atomistic simulations can more accurately address this challenge, offering direct analysis of the structural and chemical factors that govern defect stability [11,40,44–47]. For example, Cao et al. [45] used MD simulations to correlate chemical ordering at a Σ5 (210) grain boundary with a higher structural transformation temperature in face-centered cubic (FCC) CrCoNi. Notably, differences in transition temperatures were observed for weak segregation conditions (i.e., near equimolar interfacial compositions of Cr, Co, and Ni), suggesting that alloys with even stronger adsorption tendencies might exhibit a more pronounced stabilization effect. Moreover, a critical knowledge gap exists concerning how the transition from the segregated conditions at low temperatures to chemically disordered conditions at high temperatures influences the structural transition pathway. A more extensive exploration of grain boundary stability, capable of accommodating the simultaneous relaxation of interfacial structure and chemistry, is needed to probe the possible chemical configurations that emerge due to local atomic rearrangement.

In this work, atomistic simulations are used to explore the effect of segregation and local chemical order on a grain boundary transition from metastable to equilibrium states at a $[1\bar{1}0]$ symmetric tilt boundary in NbMoTaW. MD simulations in pure Nb establish a baseline of the transition in a model system without compositional complexity, revealing a structural transition with distinct low- and high-temperature grain boundary configurations. Hybrid MC/MD simulations extend the investigation to the MPEA, indicating a significantly delayed structural transformation with a transition temperature range (temperature span from the beginning to the end of the transition process) more than twice that observed in pure Nb. Structural and chemical analysis of the MPEA transition reveals localized nucleation of the equilibrium structural motifs and distinct compositional patterns between metastable and equilibrium structural motifs. These distinct segregation patterns appear as the relative enrichment of W and depletion of Nb in the equilibrium interfacial phase as compared to the metastable interfacial phase and contribute to thermodynamic stability of each phase. To decouple the relative contributions of structural and chemical relaxation, additional MD simulations are performed with random solid solution (i.e., chemically disordered) and heavily segregated chemical configurations fixed at the boundary. Surprisingly, similar transition temperature widths are observed in all chemically complex variants, suggesting that the chemical ordering state does not significantly influence the extended transition range. Instead, grain boundary segregation strongly influences the transition initiation temperature, with highly segregated interfaces only beginning to transition at very high temperatures.

## 2. Computational Methods

To investigate transition behavior in MPEAs, NbMoTaW was selected as a representative alloy due to its pronounced tendency for Nb enrichment and Mo co-segregation at low temperatures [39,48,49], coupled with a significant enthalpic drive for chemical ordering [49–51]. Additionally, the alloy's susceptibility to embrittlement by grain boundary decohesion at ambient temperatures [24,52] underscores the need for a comprehensive characterization of interface behavior across a broad temperature range. Atomic interactions were modeled using a moment tensor potential for NbMoTaW [53], which accurately captures the elemental melting points and has been extensively utilized for studying CSRO [39,48] and its effects on dislocation mechanics [53]. The same MTP was also utilized for modeling single-component Nb to enable direct comparisons of any structural transitions and to verify the effect of chemical complexity.

Further discussion of the reproducibility of the grain boundary structures with respect to other classical and machine-learning interatomic potentials is included in Section 3.1. Atomistic simulations were performed using the Large-scale Atomic/Molecular Massively Parallel Simulator (LAMMPS) software package [54], with the OVITO software [55] used for visualization and adaptive Common Neighbor Analysis (aCNA) [56] implemented for structural analysis.

MD simulations were performed on pure Nb bicrystals to investigate the grain boundary transition without segregation effects. Each bicrystal contained two Σ267 (11 11 5)[1$\bar{1}$0] symmetric tilt grain boundaries with tilt angles of $\theta = 144.4°$ (an example of grain boundary structure shown in Fig. 1(a)) and periodic boundary conditions in all directions. The exclusion of sources/sinks during the grain boundary fabrication process, such as those provided by a free surface [44,57], maintains a transformation barrier at low temperatures that traps the metastable grain boundary structure, allowing for important interfacial transitions to be studied as temperature is increased. The 0 K grain boundary configuration was obtained by rotating each grains about the [1$\bar{1}$0] axis by $\theta$/2, followed by an iterative sampling method developed by Tschopp et al. [58]. Wagih [59] recently showed that high Σ values (reflecting low coincidence in overlapping sites between adjacent lattices) can more accurately approximate the important features of segregation in polycrystalline environments as compared to low Σ (<10) grain boundaries. Furthermore, grain boundary structure and transitions in BCC metals along the same tilt axis have been extensively explored [47,60]. Each cell contained 76,368 atoms, with approximate dimensions at 0 K of $L_x = 154$ Å, $L_y = 325$ Å, and $L_z = 28$ Å. To achieve structural equilibrium, each sample was brought to a target temperature over 50 ps using the isobaric, isothermal (NPT) ensemble with an integration timestep of 1 fs. As a last step in all pure and chemically complex samples, a dynamic relaxation of the boundary structure was performed for 150ps at the target temperature. While a minimum of three independent runs were conducted at each temperature, up to ten separate simulations were performed near transition temperatures to ensure sufficient sampling of possible structural states. Single crystal samples were also simulated to obtain excess quantities.

For the MPEA, three unique chemical configurations, including a random solid solution (*RSS State*), a structurally and chemically relaxed state (*Equilibrated State*), and a fixed segregation state (*Segregated State*), were investigated to decouple the effect of simultaneous adsorption and structural relaxation on interface behavior. The RSS configuration (example shown in Fig. 1(b)) was obtained by randomly assigning atom type to Nb, Mo, Ta, and W until an equiatomic composition was achieved, followed by heating to the target temperature and isothermal annealing. The Equilibrated State (an example at 800 K is shown in Fig. 1(c)) was obtained using hybrid MC/MD methods for chemical and structural relaxation. An RSS boundary was heated first, and compositional and structural states were sampled by performing a number of trial swaps equal to 1% of all atoms for each elemental pair combination every other MD step. Each swap was accepted or rejected based on the Metropolis MC algorithm. The cell was considered chemically relaxed when the interface element concentrations changed by less than 1% over the final 500 timesteps, although chemical convergence often occurred before this very conservative cutoff. We choose to use a physical convergence criterion due to the importance of grain boundary segregation on interfacial transitions, yet the energy of the system also converges as composition stabilizes. Examples of the evolution of grain boundary composition, potential energy per atom, and the gradient of the total potential energy are shown in Figs. S1 and S2 in the Supplementary Materials. We note that, while the Equilibrated samples have experienced chemical and structural relaxation, this does not necessarily mean that the grain boundary state reaches an absolute thermodynamic equilibrium structure for a given set of conditions. The Segregated configuration was first attained from the 900 K Equilibrated state, where significant grain boundary segregation is observed, followed by annealing to the target temperature and subsequent dynamic relaxation. All samples underwent a conjugate-gradient energy minimization procedure as a final step to remove thermal noise and allow for structural analysis. Melting temperatures) for pure Nb (~2860 K) and NbMoTaW (~3485 K) were obtained using the solid-liquid coexistence method [61], allowing for homologous temperature ($T/T_m$) comparisons between the two metals. Temperatures from $0.17T_m$ to $0.95T_m$ were investigated in the MPEA samples and from $0.21T_m$ to $0.94T_m$ in the pure Nb samples.

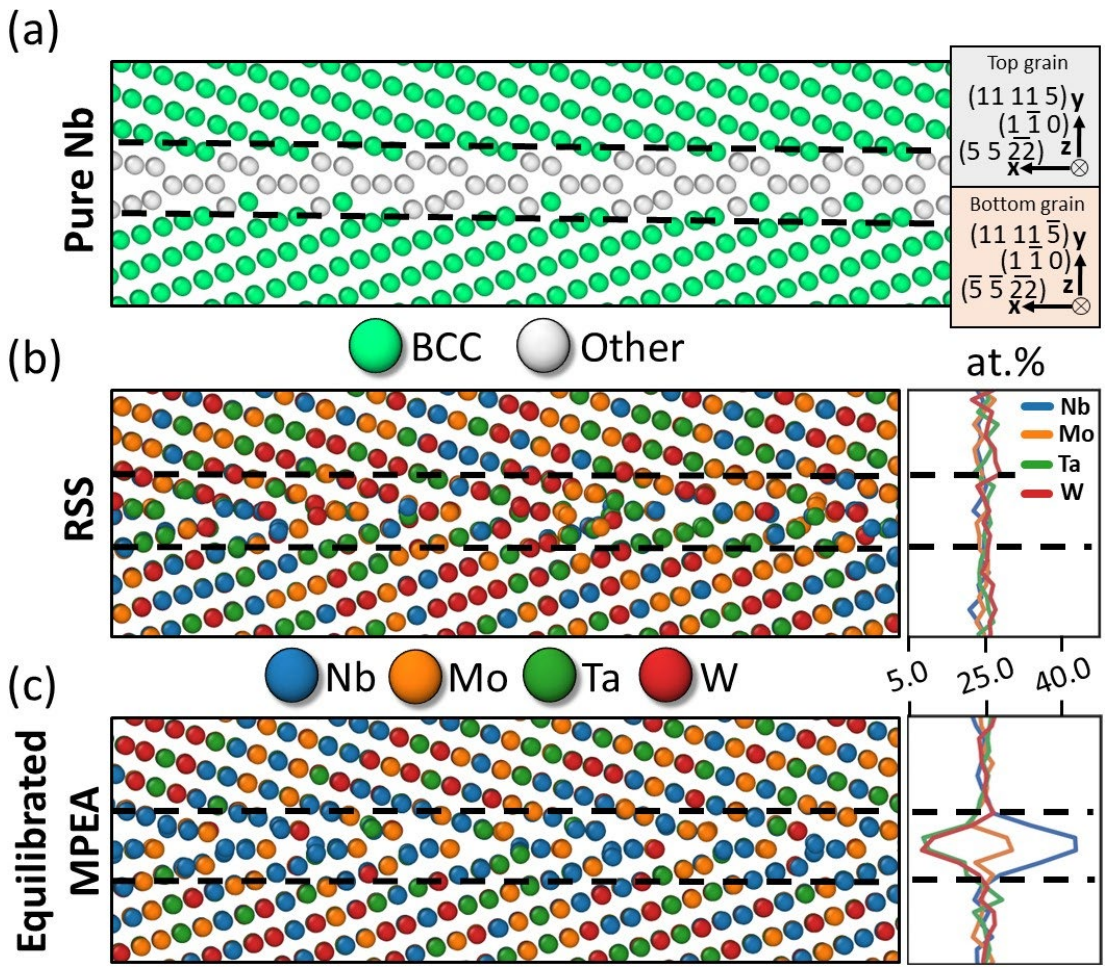

Fig. 1. Example slices of the different grain boundary states in this study, where dashed lines outline the grain boundary regions. (a) A slice from the pure Nb bicrystal with atoms colored by structure type, where green represents BCC atoms, and white represents grain boundary atoms. (b) A slice of the simulation cell with equiatomic concentrations of NbMoTaW in the RSS State colored by atom type. (c) The corresponding slice of the Equilibrated MPEA at 800 K after MC/MD simulation.

## 3. Results and Discussion

### 3.1. Frustration of the structural transition in chemically complex environments

Three common structural descriptors are used to characterize the temperature-dependency of the grain boundary structure in the pure Nb (black line) and relaxed NbMoTaW (red line), as shown in Fig. 2. Derivations of each descriptor are provided in Supplementary Note 2. Analyzing Nb first provides a benchmark for understanding the structural transition behavior in a model pure metal. In each plot, the Nb data exhibits a discontinuous change at $0.33T_m$, marking the beginning of the transition. The transition ends at $0.44T_m$, with the transition region marked with red dashed lines and shaded, covering a homologous temperature range of $\Delta T = 0.11T_m$. The evolution of the intrinsic boundary distortion is quantified and plotted in Fig. 2(a) using excess disorder, a descriptor that quantifies the structural disorder in the grain boundary relative to a single-crystal environment [62]. The excess disorder, $\Gamma_{dis}$, is obtained by first calculating a bond-orientational [63] disorder parameter, $d$, for atoms in each bicrystal and single-crystal sample at a specified temperature. This parameter measures the similarity of each atom's local bonding environment to that of its eight nearest neighbors, where a value of 0 corresponds to atoms in a perfect lattice, and a value of 1 corresponds to atoms in a liquid [62,64]. The excess disorder is then determined as:

$$\Gamma_{dis} \equiv (\Sigma_{i=1}^{N} d_i^{GB} - \Sigma_{i=1}^{N} d_i^{bulk})/2A \qquad (1)$$

where $\Sigma_{i=1}^{N} d_i^{GB}$ is the sum of atomic disorder over all $N$ atoms in the bicrystal sample, $\Sigma_{i=1}^{N} d_i^{bulk}$ is the corresponding sum of disorder over $N$ atoms in a simulation cell without defects, and $2A$ represents total grain boundary area of the simulation cell. The sudden increase of $\Gamma_{dis}$ over a narrow temperature range suggests a discrete change in boundary structure rather than general interfacial disordering, which is supported by the analysis of the grain boundary free volume (GBFV) in Fig. 2(b). Initially, the GBFV slowly increases with temperature up to $0.33T_m$, but then undergoes a rapid drop within the same transition range identified by $\Gamma_{dis}$. The decrease in GBFV is caused by a rapid reduction in average volume of grain boundary atoms relative to the average volume of bulk atoms, implying a transformation to a more compact boundary structure that reduces excess volume within the grain boundary plane. Subsequent heating to even higher temperatures resulted in increasing GBFV, which has been observed in other pure metals [65], however pre-melting to a full amorphous structure was not observed for any of the conditions studied here. Further evidence of a structural transition is found in the measurement of grain boundary thickness (Fig. 2(c)), revealing similar transition behavior as that exhibited in $\Gamma_{dis}$.

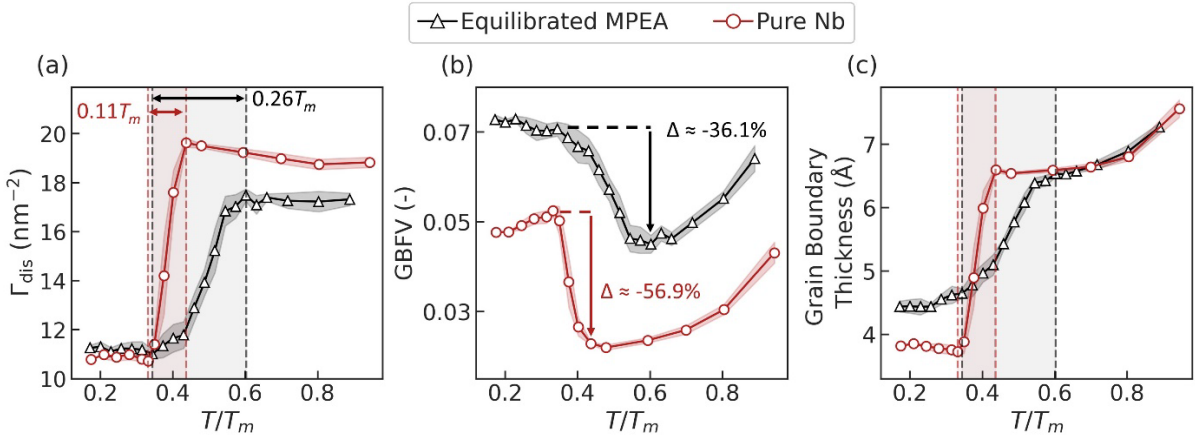

Fig. 2. The (a) excess disorder ($\Gamma_{dis}$), (b) grain boundary free volume (GBFV), and (c) grain boundary thickness as a function of homologous temperature for pure Nb and the Equilibrated MPEA. In (a) and (c), red and black shaded regions mark the identified transition regions for Nb and the MPEA, respectively. In (b), the dashed line and arrow indicate the decrease in GBFV through the transition, and values indicate the percentage decrease from the initial to final transition temperatures. All shaded bands along the trend lines in this and the following figures show the standard deviation of each measurement.

Analysis of the same interface descriptors for the Equilibrated NbMoTaW samples reveals a transformation that is frustrated compared to its pure counterpart. Similar to pure Nb at low temperatures, both the $\Gamma_{dis}$ (Fig. 2(a)) and the GBFV (Fig. 2(b)) initially remain constant with increasing temperature. In this low-temperature range, the interface composition is highly enriched in Nb, slightly enriched in Mo, and heavily depleted in both Ta and W. Site stabilization due to strong segregation tendencies can therefore play a significant role in reducing the thermodynamic driving force for nucleation and growth of a new structural phase, leading to a more gradual transformation. For example, while $\Gamma_{dis}$ in the MPEA begins to increase after $0.34T_m$, relatively slow growth up to a maximum value at $0.60T_m$ (black dashed line in Fig. 2(a)) suggests a transformation barrier not observed in the pure system. The total transition (shaded black region in Fig. 2) accounts for 26% of the homologous temperature range, or approximately 2.5 times that observed in pure Nb. Notably, the final disorder value in the MPEA is significantly lower than that observed for pure Nb at the end of the transition despite exhibiting similar values at low temperatures. Because little segregation is expected at high temperatures, this discrepancy indicates that chemical disorder also plays a key role in frustrating the structural transition. Similarly, the GBFV decreases until it reaches a minimum at $0.60T_m$, showing only a 36% decrease in GBFV over the transition (compared to a 57% decrease observed for pure Nb), providing further evidence of an incomplete atomic rearrangement into the new structural motif.

The evolution of the grain boundaries to a more compact structure should have implications for key material properties. For example, experimental studies have pinpointed grain boundaries as the primary sites for impurity-induced embrittlement [66], leading to a broad ductile-to-brittle transition temperature window (room temperature to 600 °C) [24]. This observed embrittlement has generally been attributed to weakened interfacial bonds due to oxygen embrittlement [67]. While we do not explicitly treat this phenomenon in this study, it is likely that the high grain boundary free volume in the low temperature structure creates a more hospitable environment for oxygen interstitial segregations, as compared to a more compact equilibrium grain boundary configuration. High-resolution transmission electron microscopy of nanocrystalline W suggests that non-equilibrium, high-angle interfaces contain a larger number of sites for preferential interstitial segregation [68]. Conversely, highly ordered grain boundaries, such as Σ3 coherent twins in Ni, exhibit less oxygen segregation than more random boundaries [69]. The reduction in grain boundary free volume due to interstitial impurity segregation decreases the grain boundary energy and provides a structural driving force for interfacial enrichment [70]. Although the chemical interactions of the segregating elements and the interstitial impurities should not be ignored [52], the transition to a lower energy, compact grain boundary structure should also be important and resist such embrittlement.

Differences between the Equilibrated MPEA and pure Nb are also observed in grain boundary thickness. At low temperatures, the interfacial thickness (Figs. 2(c)) in the multi-component system is initially larger than that of pure Nb, revealing how the structural disorder at the interface is moderately amplified due to chemical complexity, especially at low temperatures where compositional fluctuations can more effectively distort local bonding environments. At higher temperatures, this effect is largely reduced due to decreasing segregation and the weakened influence of chemical bonding on structural stability. Beyond $0.60T_m$, all parameters exhibit trends similar to those of the pure system.

The relaxed grain boundary structures for pure Nb from before, during and after the identified transition temperatures are shown in Fig. 3. The side view provides a parallel perspective of the interface (along the $z$-axis), including nearby BCC atoms, while the top view is tilted perpendicular to the interface (along the $y$-axis) and exclusively displays defect atoms. To differentiate between the metastable and equilibrium configurations, a descriptor that effectively distinguishes the two structures is needed. Specifically, the real component corresponding to ($l = 4, m = -1$) of the normalized spherical harmonic $\hat{Y}_m^l$ is utilized. For a complete description of this parameter, the reader is directed to Supplementary Notes 2 and 3.

Before the transition (Fig. 3(a)), the grain boundary plane consists of normal kite (labeled "N") and split-kite (labeled "S") structural units, differentiated by varied atomic densities [71,72]. During the transition (Fig. 3(b)), the coexistence of homogeneous metastable and equilibrium structures is observed, each significantly present and separated by a discrete phase junction where the blue and orange atoms meet. Black arrows denote the emergence of an "interstitial" atomic pattern in the equilibrium configuration, signifying an incommensurate number of atoms in the grain boundary plane compared to adjacent grain planes. Because the choice of interatomic potential can have significant effect on structural stability and defect evolution [73], we verify the reproducibility of the metastable and equilibrium phases using both classical and alternative machine-learning potentials. Sun et al. [74] developed a generalized embedded atom method (EAM) potential specifically for the study of grain boundary structural transitions in Nb. In studying the metastable structures of a Σ27(552)[1$\bar{1}$0] using the γ-surface method, these authors found similar normal kite and split-kite boundary phases that subsequently transition to the same equilibrium structure with interstitials after high-temperature annealing. Equilibrium structural units with the characteristic interstitial patterning were previously found in elemental BCC metals along the [1$\bar{1}$0] tilt boundary using evolutionary algorithms [47,75,76], a method that can more broadly explore the structural phase space by including operations such as insertion of atoms in the grain boundary core [72]. We further validated the equilibrated state for the tilt-angle used in this study in both pure Nb and NbMoTaW using a spectral neighbor analysis potential (SNAP) developed for NbMoTaW. The equilibrated structures obtained after annealing at 1600 K are shown in Fig. S4. The results show that both classical and machine-learning models are able to capture the important grain boundary states and simulate metastable-to-equilibrium grain boundary transitions.

To understand the energetic relationship between the two configurations and confirm the nature of the metastable-to-equilibrium transition, grain boundary energies of the complete structures were compared at 0 K, revealing that the equilibrium structure has an energy that is 5.3% lower than the metastable configuration. Therefore, once the equilibrium structure is unlocked, the system will not revert back to the metastable structure if subsequently re-equilibrated at lower temperature. To put this energy difference into context, Wei et al. [77] demonstrated that grain boundary migration in α-$Al_2O_3$ involves cooperative atomic shuffling of the grain boundary through metastable and equilibrium states that differ in energy by no more than 4.5%. Similarly, a 4% energy difference between coexisting metastable domino and equilibrium pearl phases was observed at a grain boundary in Cu [78]. Therefore, the transition reported in this study falls within the range of possible energetic states that may be sampled under different processing and thermal treatments.

After the transition, in Fig. 3(c), the equilibrium grain boundary structure is exclusively present. Importantly, the analysis of pure Nb interfacial structures reveals a highly homogenous transition between ordered states, with distinct structural coexistence regions easily discernable at intermediate temperatures.

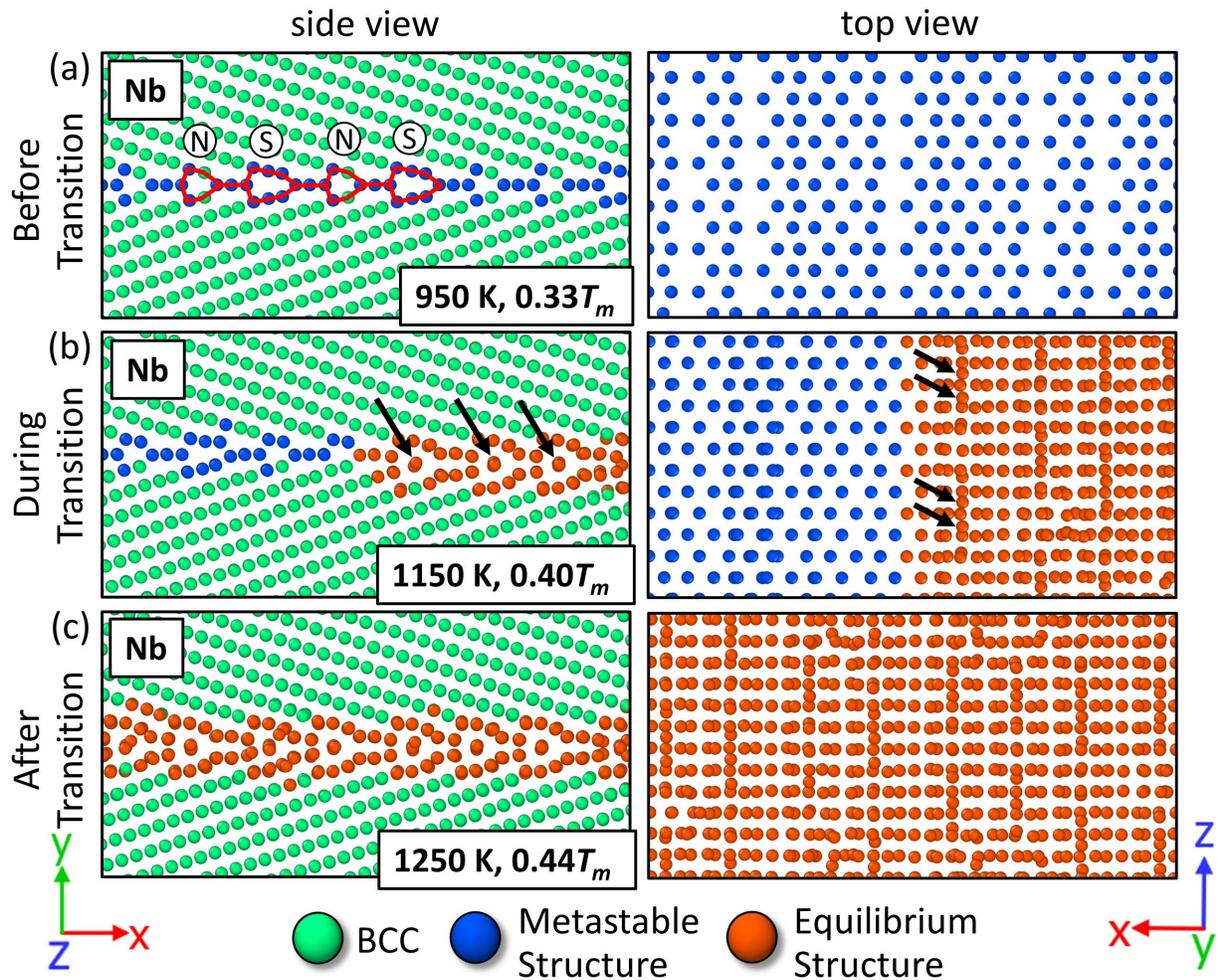

Fig. 3. Grain boundary structures for pure Nb at temperatures spanning the transition range. (a) The initial, (b) intermediate, and (c) final structures are provided to illustrate the complete structural transition and atom coloring is used to separate the BCC atoms (green) from the defect (blue and orange) atoms. Views from both **the side and top of the grain boundary are provided. In (b), arrows highlight examples of the interstitial atoms that are hallmarks of the equilibrium phase.**

The interfacial transition for the Equilibrated NbMoTaW samples is shown in Fig. 4. Before the transition (Fig. 4(a)), the boundary structure is arranged in a kite configuration similar to pure Nb, with the MPEA exhibiting a slightly larger grain boundary thickness due to defect atoms penetrating deeper into the bulk regions. Identification of the same atomic arrangement in the Equilibrated samples confirms that the metastable grain boundary state persists at lower temperatures, even after undergoing the simultaneous chemical and structural relaxation process. As the transformation takes place (Fig. 4(b)), the grain boundary structure in the MPEA transitions into a similar equilibrium structure observed in Nb. However, unlike the uniform transformation observed in the pure system, the transition in the MPEA is highly localized, resulting in disconnected regions of transformed interface. This phenomenon is highlighted in Fig. 4(b) during the transition, where the side view shows both metastable (blue atoms) and equilibrium (orange atoms) configurations overlapping within the *x-z* plane. In the top view, localized regions of both grain boundary states are observed. Even at higher temperatures (Fig. 4(c)), once the $\Gamma_{dis}$ plateaus, an incomplete transformation is observed for the MPEA, accounting for the lower magnitude of $\Gamma_{dis}$ and smaller percentage decrease in GBFV previously observed in Fig. 2.

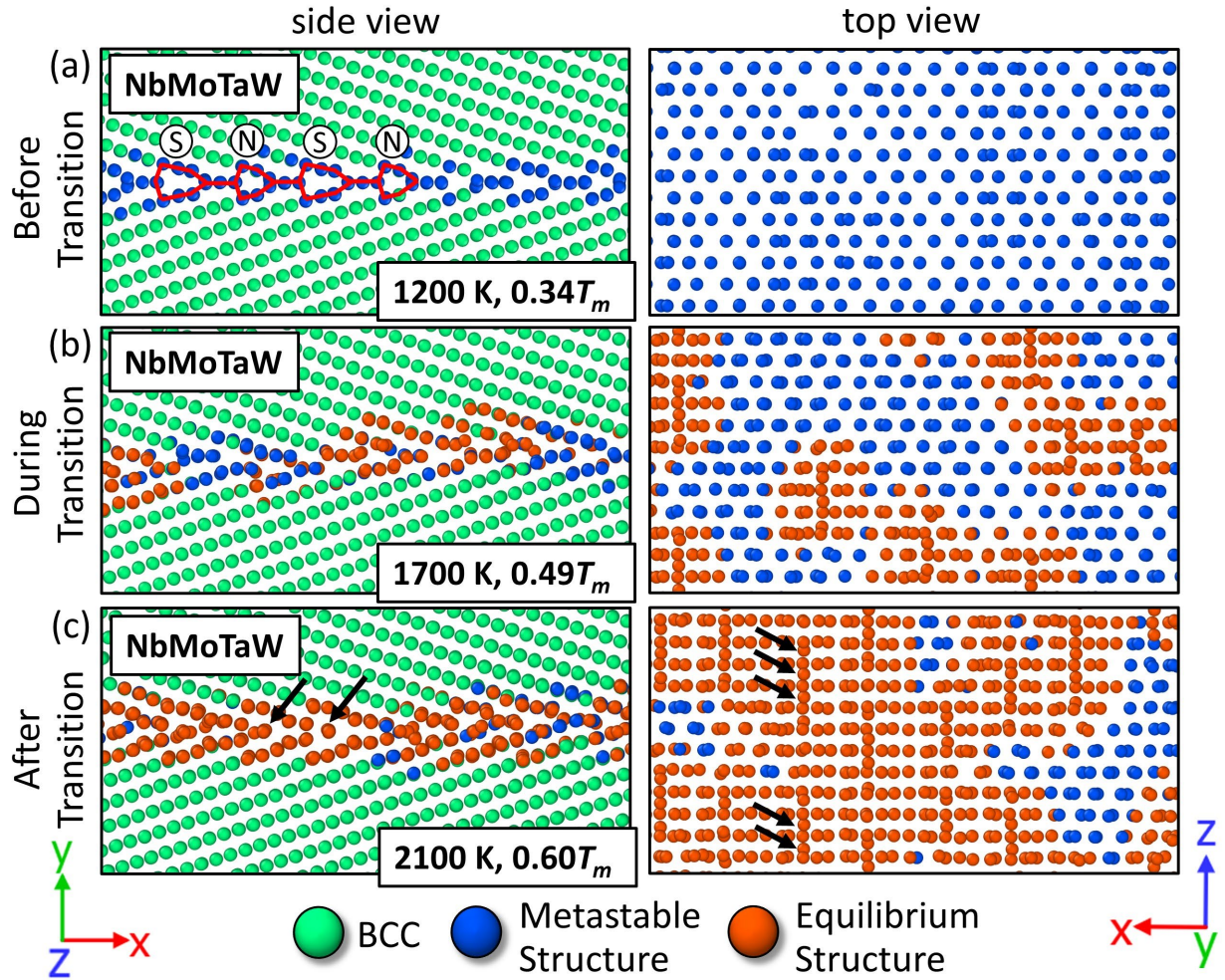

**Fig. 4.** Grain boundary structures for the Equilibrated MPEA at temperatures spanning the transition range. The initial (a), intermediate (b) and final (c) structures illustrate the incomplete structural transition. Views from both the side and top views of the grain boundary are provided and arrows indicate examples of the interstitial atoms.

To quantify the incompleteness of the transition, the fraction of the metastable state is plotted against homologous temperature for the pure Nb and Equilibrated MPEA in Fig. 5. In pure Nb, a mostly horizontal line with a metastable configuration fraction of 1.0 is observed. However, even before the transition starts, at $0.28T_m$, a subtle fluctuation occurs, indicative of a small nucleus of the transformed interface and emphasizing the metastable nature of the kite configuration. In the MPEA, this behavior is amplified due to the diverse chemical configurations that can locally modify the potential energy landscape. By $0.34T_m$ (i.e., the beginning of the transition in the Equilibrated MPEA from Fig. 2), an average of 10% of interface atoms have already transformed into the equilibrium configuration. This transformation likely occurs at interface regions with energetically unfavorable local compositions, due to desorption of strongly segregating elements, resulting in destabilization of the local structure. Conversely, at elevated temperatures, the completeness of the transformation is restricted by either residual segregation, entropy-driven chemical disorder, or a combination of these two effects. A nearly complete transformation is observed for pure Nb at the end of the transition (less than 1% of atoms are associated with the metastable configuration), while 8% of the boundary remains in the metastable state at $0.60T_m$ in the MPEA.

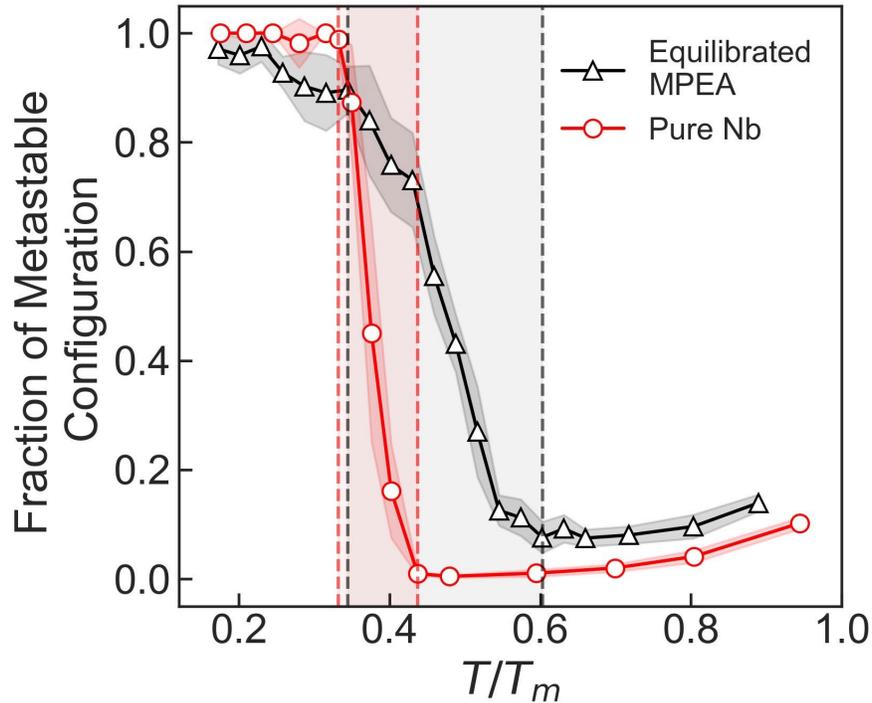

**Fig. 5. The fraction of the metastable grain boundary structure as a function of homologous temperature for the Equilibrated MPEA (black) and pure Nb (red) systems. The dashed lines and shaded regions indicate the previously identified transition temperature ranges.**

The dramatic morphological difference of the metastable and equilibrium structural states in the MPEA implies that grain boundary properties will also be significantly altered. The frustrated transition in the MPEA over a wide temperature window will result in a grain boundary that behaves non-uniformly from intermediate to high temperatures [79] as well as the delay of key grain boundary behaviors. For example, one can expect variations in structurally dependent processes, such as dislocation transmission and grain boundary sliding. Tucker at al. [80] observed a correlation between higher interfacial excess free volume and reduced resistance to grain boundary sliding and migration in pure Cu and Al, attributed to enhanced local atomic shuffling under lower shear stresses. Therefore, grain boundary sliding could occur more readily in the MPEA at lower temperatures, as a transition to the more compact structure with lower excess free volume is significantly delayed. Similar dynamic mechanisms are likely also influenced by the chemical inhomogeneity in MPEAs, requiring further investigation to elucidate the role of both structure and chemistry on grain boundary properties.

## 3.2. Segregation state and chemical ordering in the Equilibrated MPEA

Next, the temperature-dependent segregation states in the Equilibrated MPEA are identified to understand the role of dynamic chemical environments in frustrating the transition. Fig. 6 provides structural and compositional comparisons between Nb and the Equilibrated MPEA at a very low temperature ($0.14T_m$) in order to illustrate both the starting structure and the initial segregation pattern in the alloy. Black lines serve as reference for the repeating structural units identified in Figs. 3 and 4 in both Nb and the MPEA. In Figs. 6(a) and (b), coordination number and hydrostatic stress for grain boundary sites are shown in pure Nb, demonstrating how important local descriptors vary spatially. Arrows highlight the correlation between over-coordination and highly compressed sites in the elemental interface. In the MPEA, Mo atoms dominate filling of these sites, resulting in a periodic segregation pattern, as indicated by arrows in Fig. 6(c). Additional Mo decoration is also observed at sites with slightly reduced compressive stresses just outside the grain boundary core (blue in Fig. 6(a)), resulting in a partial bilayer configuration, while Nb occupies the remaining grain boundary sites. The enrichment of Nb and Mo can be partly attributed to the large difference in melting points between the pure base elements (Nb < Mo << Ta < W). Because melting point is positively correlated with bond energy [81], Nb and Mo incur the smallest bond energy penalty in under- and over-coordinated sites. Furthermore, differences between Nb and Mo site preferences are related to the elastic strain relaxation induced by smaller Mo atoms segregating to compressed grain boundary sites, and vice versa for Nb [39]. With complex chemical interactions largely minimized due to Ta and W depletion, the primary driving force for stabilizing the metastable grain boundary structure is the reduction of interatomic distortion through mechanical relaxation. We note that increasing temperature leads to a reduction in the segregation tendency and, therefore, the dominant site occupancies shown in Fig. 6.

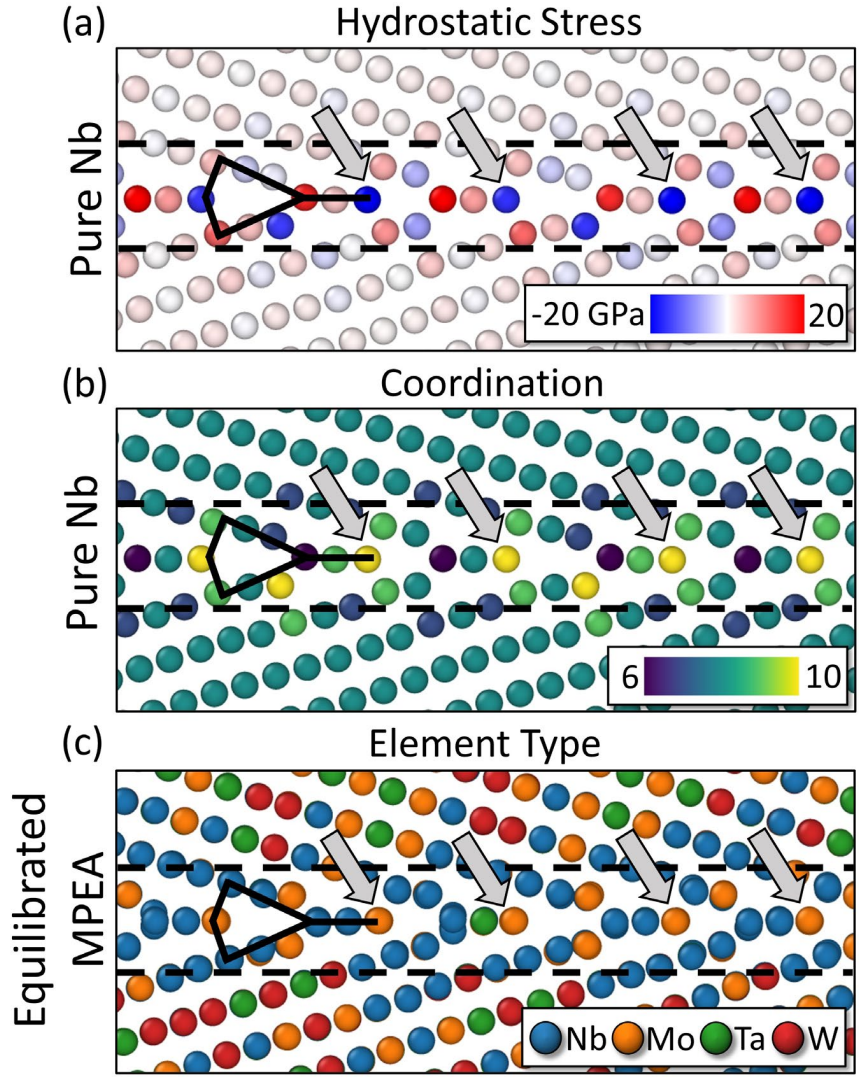

**Fig. 6. Grain boundary configurations in the pure Nb and Equilibrated MPEA models at 0.14$T_m$, colored according to (a) hydrostatic stress, (b) coordination number, and (c) elemental type. The structural unit pattern is identified using black lines in (a-c) to show how grain boundary structure correlates with elemental occupancy. Arrows highlight the correlation between compressive stresses and over-coordinated sites with Mo segregation. Dashed lines in each plot are used to indicate the grain boundary region.**

In Fig. 7, grain boundary concentrations for each element are plotted as a function of temperature. The vertical dashed lines designate the transition temperature range for the Equilibrated MPEA and the horizontal dotted line indicates equiatomic fractions (i.e., 0.25 of each element). Consistent with the interfacial analysis in Fig. 6, the boundary composition at $0.14T_m$ (i.e., 500 K) is predominantly composed of Nb and Mo atoms, accounting for nearly 94% of grain boundary atoms. With increasing temperature, enrichment of Nb decreases and the depletion of Ta and W tends towards equiatomic concentrations. Despite these trends, enrichment of Nb and depletion of W remain throughout the transition, possibly contributing to the localized stabilization of the metastable configuration into high temperatures. Interestingly, the concentration of Mo at the grain boundary does not change much with temperature, remaining slightly enriched across the entire temperature range.

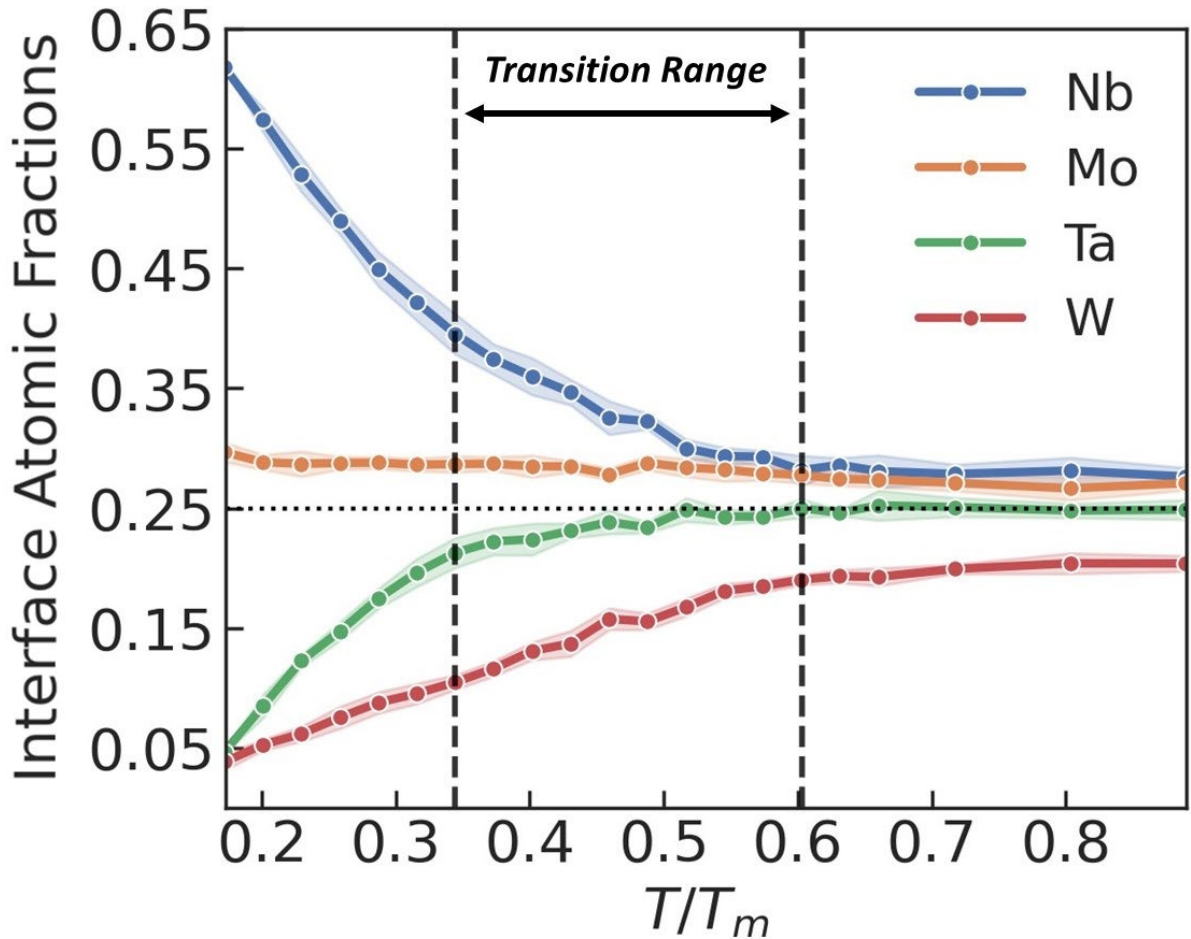

**Fig. 7.** Element concentrations at the grain boundary as a function of homologous temperature. Vertical dashed lines at $0.34T_m$ and $0.60T_m$ indicate the identified transition range for the MPEA. The horizontal dotted line shows the equiatomic concentrations.

During the transition, the reduction of GBFV could be expected to result in significant changes in the variety of local structural environments at the interface. Evidence of structure-composition relationships at grain boundaries in chemically complex environments [39,40,42] suggests that segregation tendencies are highly susceptible to such local changes. Despite this, Fig. 7 does not capture any abrupt compositional variations, likely due to the slow growth of the equilibrium phase which smooths out the differences in chemical patterns over a broad temperature range. For a detailed understanding of the compositional differences between the two phases, local concentrations in each grain boundary structure type are calculated and plotted for each element in Fig. 8. Elemental concentrations for atoms in the respective metastable and equilibrium structures are calculated using the same normalized spherical harmonic descriptor implemented for structural identification in Figs. 3 and 4. Solid lines denote the concentration of each element in the metastable grain boundary structure and dashed lines indicate the corresponding concentration in the equilibrium structure. Our focus is on temperatures where sufficient amounts of each phase are present (>25%) to ensure statistical significance, limiting the analysis to temperatures between $0.40T_m$ and $0.52T_m$. Distinct compositional differences are indeed observed between the two phases. For example, the concentration of Nb in the equilibrium structure is ~6 at.% lower than the metastable structure for all temperatures shown here (Fig. 8(a)). In contrast, Mo and W become more enriched in the equilibrium structure, with this effect most notable for W where an increase of ~5 at.% is observed (Fig. 8(d)). Ta exhibits little concentration change induced by the transition (Fig. 8(c)).

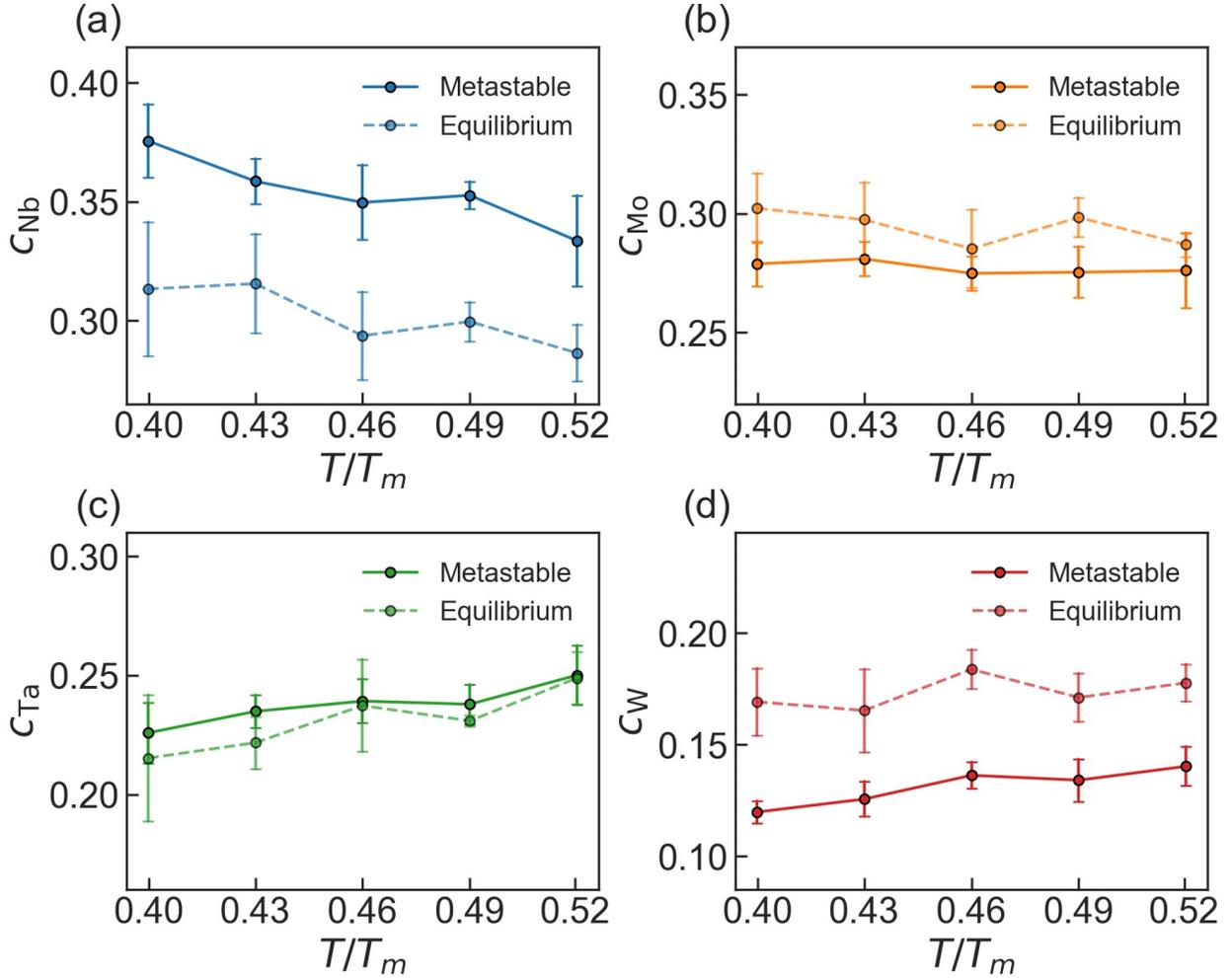

**Fig. 8. Grain boundary concentrations for each element in the Equilibrated MPEA interface, where the solid and dashed lines indicate the local concentrations in the metastable and equilibrium structures, respectively.**

To provide a physical interpretation of these segregation patterns, the local structural environment for each element in each phase is evaluated using the averaged first nearest-neighbor coordination number for grain boundary atoms (Fig. 9). As shown previously, the coordination number is highly sensitive to the local atomic configuration and is often used to correlate elemental properties and segregation tendencies. For example, Geiger et al. [39] showed that crowding by atomic neighbors in over-coordinated grain boundary sites results in compressed grain boundary sites that attract solutes with intrinsically smaller atomic volumes, while under-coordinated sites have more room to accommodate larger ones. This trend is captured in Fig. 9, with the relatively larger Nb and Ta atoms exhibiting lower average coordination than the smaller Mo and W atoms in both structures. During the structural transition, an increase in average coordination number is expected due to the reduction in grain boundary free volume that rearranges grain boundary sites into a more compact bonding configuration. Despite this expected change, negligible differences in average coordination number are observed for Nb (Fig. 9(a)) and Ta (Fig 9(c)) between the two phases. This result, combined with the decrease in Nb concentrations from the metastable to the equilibrium structure, suggests that the number density of preferred local environments (i.e., large free volume sites) for Nb atoms is decreasing through the transition temperature range, in combination with temperature-induced desorption. On the other hand, there is a significant increase in coordination number for Mo and W between the two phases. This trend, coupled with the increase in Mo and W concentrations from the metastable to equilibrium structure (Figs. 8(b) and (d)), suggests that the change in local bonding environment promotes increased concentrations of these two elements. Interestingly, the equilibrium structure is stabilized by only a modest increase in W (~4-5%) and Mo (~1-3%) concentrations compared to the average grain boundary concentration, as shown by the difference between equilibrium and metastable concentrations in Figs. 8(b) and (d). Despite this, the transition remains highly frustrated due to the strong Nb segregation character that locally stabilizes the metastable structure.

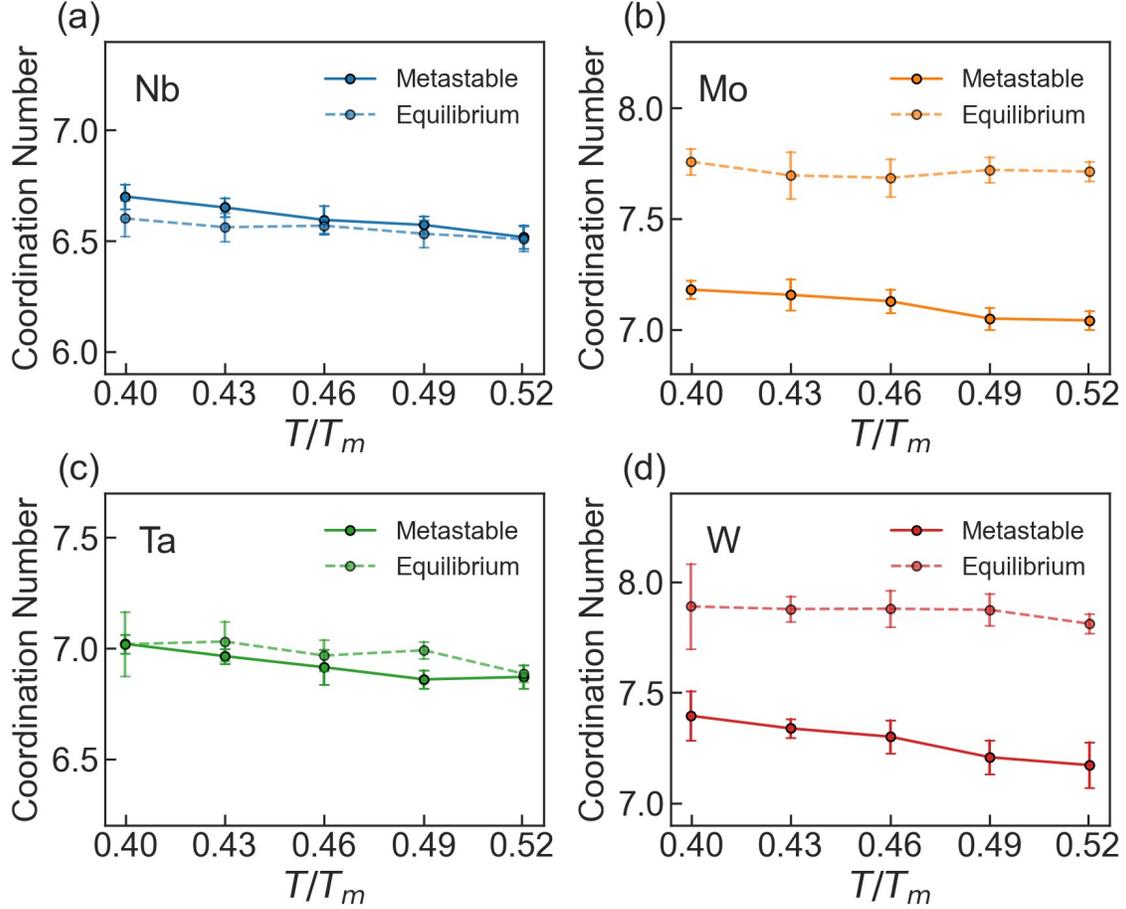

**Fig. 9. Average first nearest-neighbor coordination number for atoms in the Equilibrated MPEA grain boundary with the solid and dashed lines indicating the concentrations in the metastable and equilibrium phases, respectively.**

It is worth noting that the chemical interactions, both within the grain and the grain boundaries, may play a role in the temperature-dependence of the transition. To evaluate the pairwise interactions among different elemental pairs in bulk and grain boundary environments, the first-nearest neighbor Warren-Cowley order parameter [82] $\alpha_1^{ij}$ was used for atomic subsets contained within the two structures (Fig. S5). A detailed description of this order parameter is included in Supplementary Note 4. At lower temperatures, the dominant role of mixing enthalpy leads to greater pair interaction probabilities for certain elements (e.g., Mo-Ta, Nb-Mo in Fig.

S5(b)), resulting in negative $\alpha_1^{ij}$ values, while positive $\alpha_1^{ij}$ values reflect repulsive pair interactions (e.g., Ta-Ta, Nb-Ta in Figs. S5(a) and (b)). The preferential bonding states are likely related to differences in atomic size and electronegativities between the larger Group 5 (Ta, Nb) and smaller Group 6 (Mo, W) elements [83], approaching pseudo-binary type ordering at very low temperatures dominated by B2 Mo-Ta interactions. As temperature increases, the drive to form energetically favorable bonds is overcome by increasing configurational entropy and long-range chemical frustration, leading to an order-to-disorder transition. The disordering behavior occurs gradually, complicating the identification of a critical transition point. However, between 1100 and 1300 K (0.31 and 0.37$T_m$), the regression towards zero (i.e., a random solid solution) becomes approximately linear and differences in chemical ordering become very small. This transition point aligns with previous reported values when considering only short-range interactions such as those in the first-nearest neighbor shell [50,83]. A more complete representation of this critical temperature will include longer-range interactions as well as low temperature calculations [50,84]. At the grain boundary, different bonding relationships emerge as a result of elemental segregation and depletion. However, heavy Nb segregation conflates the effects of chemical ordering and structural relaxation on the grain boundary stability at lower temperatures. This is captured by the very negative $\alpha_1^{Nb-Mo}$ interactions in grain boundary sites (Fig. S5(d)). Because the Nb and Mo unambiguously segregate to distinct sites in correlation with structural relaxation, the co-segregation behavior or chemical effect is obscured. Moreover, the previously identified order-to-disorder temperature indicates that chemical ordering should play a secondary role to segregation-induced structural relaxation due to the high structural transition temperature. Nonetheless, a significant consequence of a more compact grain boundary structure (i.e., more sites with closer neighbors) is the increased number of chemical interactions between solutes. The enhanced interaction between solutes can potentially lower the energy barrier for transformation if the grain boundary transition leads to a greater number of attractive neighbors or raise the transformation barrier when unfavorable neighbor interactions are the result. In the latter case, higher temperatures for transformation would likely be required. This effect should become more pronounced in alloys with higher order-disorder transition temperatures, allowing chemical ordering to persist beyond the segregation regime. For example, the addition of V to NbMoTaW can drastically increase the disordering temperature due to strong bonding of V with other elements [85], indicating a more clear role of chemical order on grain boundary stability.

### 3.3. Impact of segregation state and sluggish diffusion on the transition

In the following section, the effect of local segregation state on grain boundary stability is probed using a random solid solution (RSS) configuration and a highly segregated configuration extracted from MC/MD simulations at the low temperature of 900 K (Segregated). The RSS interface comprises randomly placed atoms with elemental concentrations of approximately around 25%, representing a chemically disordered interface. On the other hand, the Segregated interface has significant elemental enrichment, with a composition of 50.4 at.% Nb, 28 at.% Mo, 13.5 at.% Ta, and 8.2 at.% W. To isolate the structural relaxation without any additional chemical relaxation, only MD simulations were conducted for each boundary type at various annealing temperatures (i.e., no MC steps were executed, and the grain boundary composition remained constant).

In Fig. 10(a), the evolution of $\Gamma_{dis}$ is presented for each of the two new interfacial states, alongside the previous findings from the Equilibrated MPEA. The transition widths for both the RSS ($0.26T_m$) and Segregated ($0.23T_m$) conditions cover a similar fraction of the total homologous range as that of the Equilibrated MPEA ($0.26T_m$). However, differences in the initiation temperature for each system demonstrate the influence of segregation state on the transformation. For example, in the RSS configuration, the transition starts and ends at temperatures approximately 200 K lower than those for the Equilibrated MPEA, occurring at $0.29T_m$ and $0.55T_m$, respectively. In stark contrast, the beginning of the structural transition for the Segregated MPEA is considerably delayed to higher temperatures. For the strongly segregating grain boundary, the $\Gamma_{dis}$ decreases slightly in the temperature range from $0.26T_m$ (900 K) to $0.57T_m$ (2000 K), indicating that the overall disorder in the bulk regions actually increases faster with temperature than the disorder at the interface. Subsequently, after $0.63T_m$ (2200 K), the $\Gamma_{dis}$ increases monotonously to the last investigated temperature at $0.95T_m$. The continuous increase in $\Gamma_{dis}$ complicates the precise determination of the transition end in the Segregated MPEA, yet a change in the slope of this curve near $0.83T_m$ (2900 K) suggests that the transition reaches a point of stagnancy before the interface continues to disorder due to high temperature.

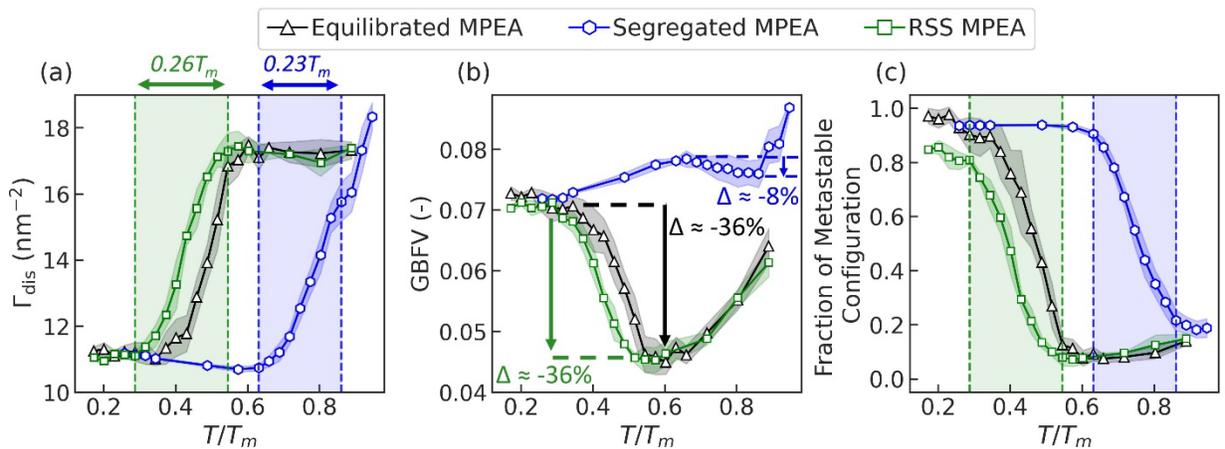

**Fig. 10. (a) Excess disorder ($\Gamma_{dis}$) (b) grain boundary free volume (GBFV) and (c) fraction of the metastable configuration of the Segregated, Random Solid Solution (RSS), and Equilibrated chemical configurations as a function of homologous temperature. Colored dashed lines indicate the transition ranges for each interfacial ordering condition.**

To gain deeper insights into the transition for the RSS and Segregated MPEAs, Figs. 10(b) and 10(c) present the GBFV and fraction of the metastable configuration, respectively. The Equilibrated and RSS MPEA exhibit a comparable decrease in GBFV of approximately 36%, reflecting similar accommodation of atomic rearrangement into a more densely packed structure. However, Fig. 10(c) reveals a greater fraction of transformed interface at the lowest homologous temperature for the RSS as compared to the Equilibrated MPEA. This suggests greater susceptibility to local transitions in the chemically disordered state, likely in high-energy regions with sub-optimal elemental occupancy. In contrast, the Segregated MPEA shows a more modest decrease in GBFV of only around 8%. While this may lead one to believe that the Segregated MPEA undergoes a much more restricted transformation response, analysis of the fraction of the metastable configuration shows that significant interfacial reconstruction is achieved, concluding with ~78% of the interface consisting of the equilibrium configuration. The delayed transition to significantly higher temperatures in the Segregated MPEA results in competing dynamics between GBFV decrease due to the transition and GBFV increase induced by increasing temperature, resulting in an overall dampened response and reduced net GBFV decrease during the transition. Importantly, the sharp upturn in GBFV after $0.83T_m$ aligns with the slight change in slope at the end of the transition in Fig. 10(a) and is similar to the responses of the both the RSS and Equilibrated MPEAs after their respective transition end points.

In Fig. 11, boundary structures for the RSS and Segregated MPEA samples are shown at both before (Figs. 11 (a) and (c)) and after (Figs. 11(b) and (d)) their transitions. Despite their significantly different transition start temperatures, the boundary structures for each ordering condition just before the transition are the same, being comprised of kite-type structural units. At the end of the transition, both MPEAs undergo substantial structural transformations, evident in the top view perspective by the density of orange atoms which represent the equilibrium structure. Comparisons of the two transformed interfaces in Figs. 11(b) and (d) show more blue atoms in the Segregated MPEA, reflecting the greater fraction of residual metastable configuration at the end of the transition for the chemically-ordered state (0.22) as compared to the RSS (0.08). The high-temperature structural frustration in the Segregated MPEA, coupled with the significantly delayed transition start temperature, confirms the influence of segregation in delaying the metastable-to-equilibrium transition in the Equilibrated MPEA. However, due to restricted desorption, the effect of segregation is more pronounced in the Segregated state.

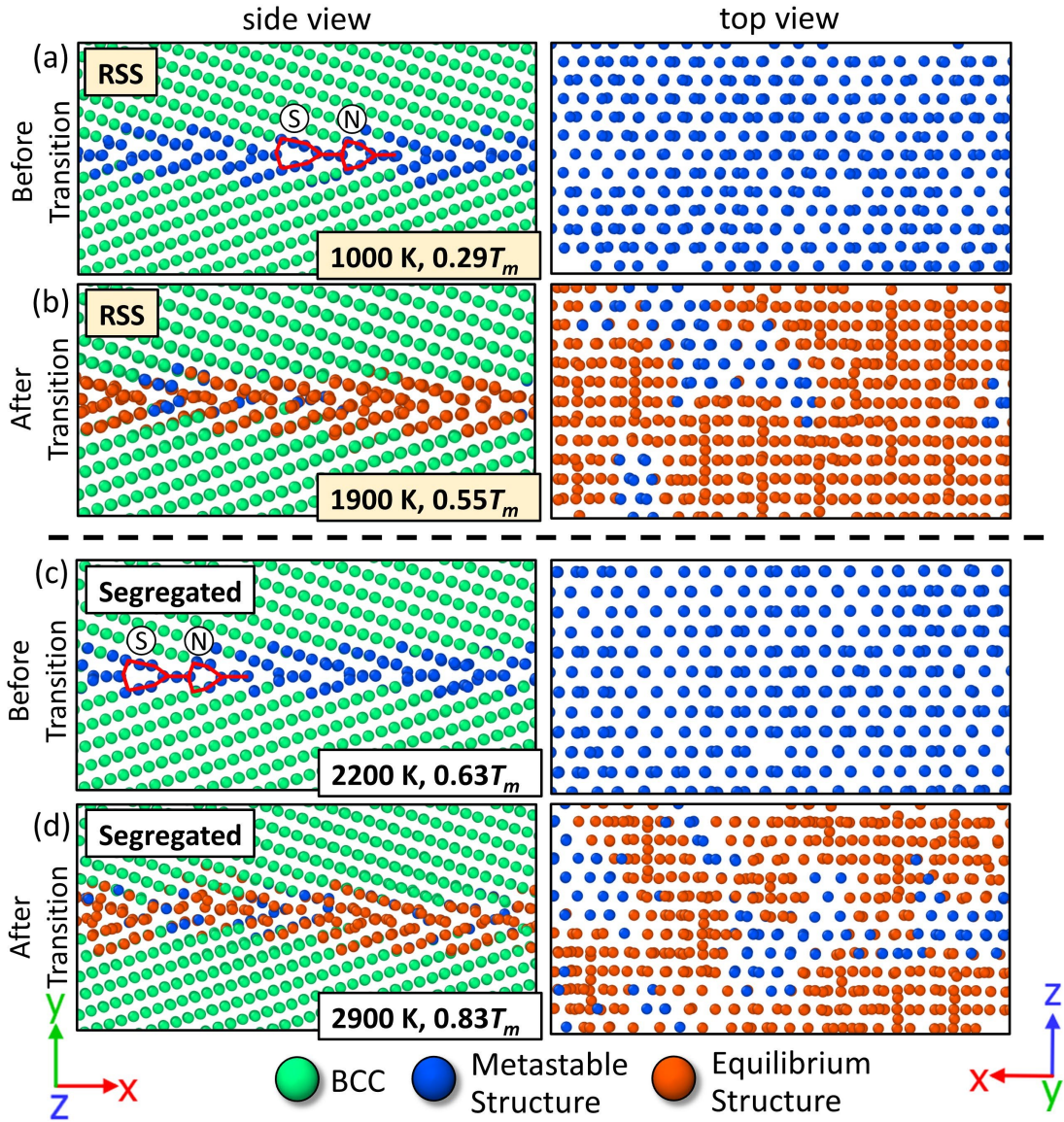

**Fig. 11.** Grain boundary structures for RSS and Segregated MPEA configurations at temperatures associated with the beginning (a, c) and end (b, d) of their respective transitions.

Since segregation state can alter the mobility of interface atoms and therefore the capacity to rearrange into a new structure, grain boundary diffusivity measurements were conducted to assess kinetic contributions to the frustrated transition. First, the average mean-square displacement $\langle R^2 \rangle$ (MSD) of grain boundary atoms in the Segregated and RSS states was calculated at three temperatures in 100 K increment below the transition initiation temperature.

To account for slight structural fluctuations, the MSD was calculated for all atoms within a 1.0 nm wide region centered at the mean *y* position of defect atoms within each grain boundary. For the RSS state, this approach requires diffusion simulations near and below ~$0.26T_m$, a temperature far less than the melting point, and thus impractical to implement due to the very long MD run times needed to acquire sufficient atomic hop statistics. For example, during five separate diffusion simulations at the lowest temperature investigated (i.e., $0.20T_m$), only a few scattered atomic jumps were observed for grain boundary atoms over a 15 ns run time (Fig. S6) and a diffusivity could not be reliably measured. To enable appreciable atomic motion on MD timescales, 1 at.% vacancies (~20 vacancies) were introduced into the grain boundary and three simulations were performed for each chemical configuration and temperature, each with different random vacancy placements. By introducing excess vacancies, grain boundary diffusion is sped up yet inherently biased towards vacancy diffusion mechanisms. Recent studies have highlighted the importance of self-interstitial atoms and exchange mechanisms in facilitating diffusion, especially at the grain boundaries [57,86,87]. In this study, the addition of vacancies is done in the same manner across all samples allowing for self-consistent comparisons of atomic mobility in the two chemical environments. Importantly, the other diffusion mechanisms are not explicitly restricted and may also have an effect on the diffusion process. A comparison set of simulations was also performed at 900 K without additional vacancies, to directly measure atomic mobility without any confounding factors. The diffusion coefficient for each segregation state and temperature was determined from the linearized relation between MSD and time following an initial relaxation period [88].

Fig. 12(a) illustrates the MSD values for the Segregated (blue) and RSS (green) grain boundary atoms as a function of time at $0.26T_m$ (or 900 K). Additional MSD data for various temperatures is provided in Fig. S7. The RSS boundary exhibits a much faster increase in MSD compared to the boundary with significant segregation, indicating a more rugged potential energy landscape in the Segregated state leading to reduced and localized diffusion [89]. In contrast, without chemical relaxation, grain boundary sites will exhibit lower migration barriers, thus promoting a higher rate of atomic hops and reducing the barrier for initiation of structural transformation. This disparity in atomic mobility provides an additional factor, beyond higher relative grain boundary energy, to explain why the RSS boundary begins to transition at lower temperatures. Since the atoms in the RSS are more likely to hop to new sites, they exhibit a greater propensity to accommodate the driving force for transformation. This also suggests that the phenomenon of 'sluggish diffusion' in MPEAs is more pronounced when heavy segregation is present than in the case of maximized configurational entropy (i.e., the RSS state).

In Fig. 12(b), the Arrhenius relation between diffusivity and reciprocal temperature shows a clear difference in diffusion kinetics between the segregation conditions. Dashed lines represent the fitted relationship between the measured diffusivities. The diffusivity error is calculated from the standard deviation of the MSD vs. time relation for all grain boundaries at a given temperature and chemical state. The activation energy, $Q$, is calculated from the slope of the respective lines. An important observation is the significantly higher diffusivity of the RSS boundary compared to the Segregated boundary. Even at 700 K, the RSS configuration exhibits larger diffusivity than the Segregated configuration at 1000 K. This difference, along with the lower activation energy (38 kJ/mol for RSS vs. 49 kJ/mol for Segregated), confirms that the atoms in the chemically random interface are able to migrate more easily, contributing to a lower transition temperature. However, this behavior alone cannot fully explain the drastic differences in initiation temperature. Extrapolating the Arrhenius relations to the respective transition start temperatures reveals the diffusivity for the RSS configuration near the transition is ~5.1 × 10$^{-14}$ m$^2$/s, while for the Segregated configuration near the transition, it's an order of magnitude higher at ~2.5 × 10$^{-13}$ m$^2$/s. This difference indicates that, although diffusion is restricted, the dominating factor in the delayed metastable-to-equilibrium transition is segregation-induced structural effects rather than restricted kinetics. For the simulations without extra vacancies, the grain boundary diffusivity for the RSS state was 1.97 × 10$^{-14}$ m$^2$/s while the Segregated state did not exhibit discernible atomic motion (Fig. S8). Therefore, the overall greater diffusivity of the RSS state relative to the Segregated state is a natural consequence of the segregation state, with or without excess vacancy concentrations.

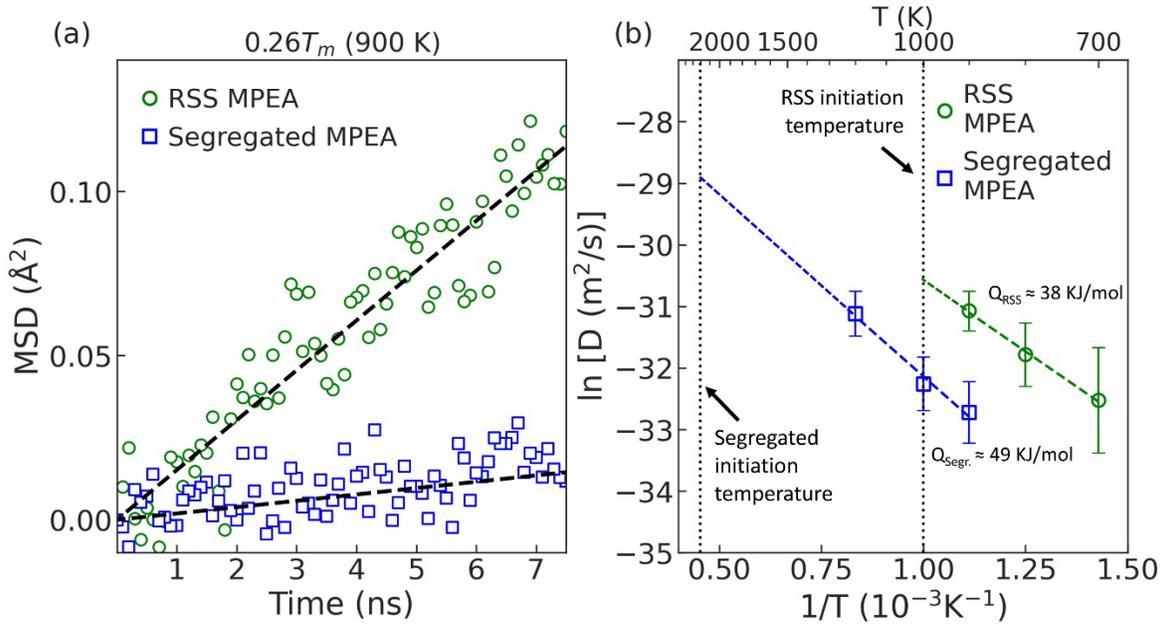

**Fig. 12. (a) Mean-squared displacement of grain boundary atoms at 900 K as a function of time for the RSS (green circles) and Segregated (blue squares) chemical configurations. (b) Arrhenius plot of grain boundary diffusivity for the RSS (green circles) and Segregated (blue squares) states. Linear fits according to the respective data points for each ordering configuration are used to calculate the diffusivity values at the beginning of each transition (dotted vertical black lines). The error at each point in (b) is the standard deviation of the slopes from the MSD versus time relationship acquired for each grain boundary at a specific temperature.**

## 4. Summary and Conclusions

In this study, atomistic simulations were used to explore the effect of segregation and chemical complexity on a metastable-to-equilibrium grain boundary transition in an MPEA. The results presented here lead to the following conclusions:

(1) Structural transitions in MPEAs can be highly frustrated, spanning a broader temperature range compared to a pure reference system. Unlike the rapid nucleation and growth of the new grain boundary structure observed in pure Nb, the transition in the Equilibrated MPEA is spatially localized, resulting in scattered regions of untransformed interface, even at high temperatures.

(2) Heavy segregation of Nb decreases with increasing temperature in the Equilibrated MPEA, destabilizing the metastable grain boundary structure and initiating the transition. At temperatures with structural coexistence, segregation patterning emerges, characterized by a relative decrease in Nb concentration and an increase in W concentration in the equilibrium motifs. This change in composition correlates with an increase in the number density of preferred local structural environments for the smaller elements.

(3) Analysis of different segregation states shows that the transition range is minimally affected by local chemical enrichment at the MPEA interface. The transition range for all MPEA variations was approximately 2.5 times that observed in the pure system, emphasizing the predominant influence of general chemical complexity on transition temperature width.

(4) In contrast, the initiation temperature of the transition is strongly influenced by the segregation state. Increased segregation (RSS → Equilibrated → Segregated) leads to higher transition start temperatures. Analysis of grain boundary kinetics suggests that thermodynamic stabilization induced by segregation significantly contributes to delaying the initiation temperature, rather than only being the result of sluggish kinetics.

Overall, this study provides fundamental insight into the thermodynamic and kinetic driving forces underlying interfacial structural transitions in MPEAs. Future experiments should account for the long-range diffusion required to obtain equilibrium structural and compositional patterning, as it will inevitably affect the transition rate and probability. Our findings suggest that significant chemical segregation is not a prerequisite for the delayed transition and its associated structural features. This further supports the motivation for in-situ electron microscopy experiments, even in non-equilibrium conditions, to better understand the role of chemical complexity and ordering on structural transitions.

**Acknowledgments**

This research was primarily supported by the National Science Foundation Materials Research Science and Engineering Center program through the UC Irvine Center for Complex and Active Materials (DMR-2011967). The authors thank Prof. Shyue Ping Ong for providing the machine learning interatomic potential used in this work.

# Supplementary Materials

**Supplementary Note 1**

During the hybrid MC/MD simulation, the interface concentrations were monitored to assess the chemical convergence of the cell. Alternatively, the potential energy per atom could be used to determine an adequate convergence point, as shown in Fig. S1(a) for a sample a 600 K. In the final ps of the MC/MD procedure, the potential energy becomes relatively flat indicating that the equilibration process converges to a constant value. Comparisons of the interface concentrations (Fig. S1(b)) over the last ps confirm that the grain boundary composition is no longer exhibiting large fluctuations. Fig. S2 shows the gradient of the total potential energy and moving averages calculated with a window size of 50 MD timesteps (out of 2000 MD timesteps, or 2.5% of total datapoints). The convergence of the gradient of potential energy occurs around the 7 ps mark, at which point the moving average fluctuates around a constant value indicated by the dashed line. The choice of 8 ps for total MC/MD procedure is thus appropriate. It should also be noted that convergence occurs more quickly at higher temperatures than at 600 K (the lowest temperature simulated), although all simulations run for the same number of timesteps regardless of temperature.

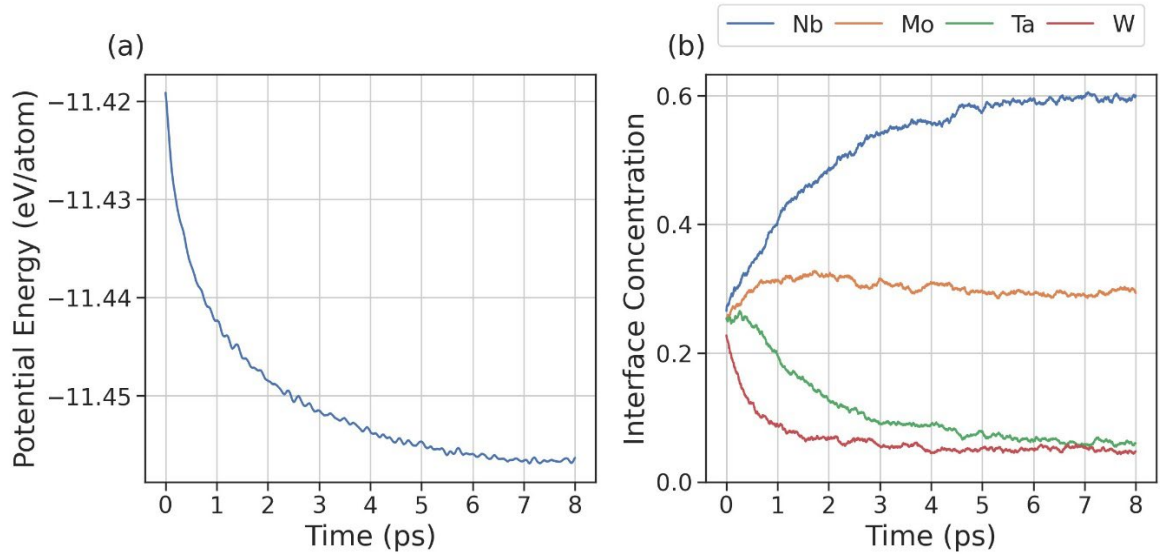

Fig. S1. Different metrics for determining stability of the system for a sample equilibrated at 600 K. (a) The evolution of the potential energy per atom (eV/atom) and (b) the change in interface concentrations as a function of time during the MC/MD procedure. In (b), each color corresponds to the respective elemental concentration at the grain boundary.

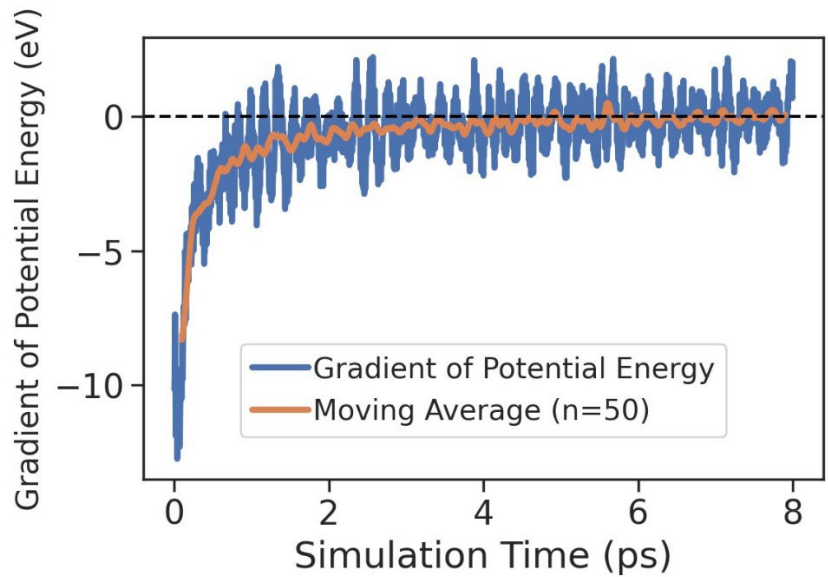

Fig. S2. The gradient of potential energy (blue) during the MC/MD procedure at 600 K, and the corresponding moving averages with a window interval of 50 (orange).

**Supplementary Note 2**

The $\Gamma_{dis}$ is obtained by first calculating a bond-orientational [1] disorder parameter, $d$, for atoms in each bicrystal and single-crystal sample. This parameter measures the similarity of each atom's local bonding environment to that of its eight nearest neighbors and is obtained utilizing a method outlined by von Alfthan et. al. [2]. First, a geometric bond relationship between the central atom, $i$, and its neighboring atoms, represented by the set $N_i$, is established. In this work, we specifically consider the 8 nearest neighbors as the set $N_i$, although the choice of neighboring atoms can also be determined by fixing an interaction radius around the central atom. This selection adequately represents the local disorder in both crystalline and defect regions. The local structure around particle $i$ is defined as a set of complex numbers:

$$\bar{Y}_m^l(i) = \frac{1}{N_i} \sum_{j \in N_i} Y_m^l(\theta_{ij}, \Phi_{ij}) \qquad (1)$$

where $\theta_{ij}$ denotes the colatitude and $\Phi_{ij}$ represents the azimuthal angle betweens atoms $i$ and $j$. The normalized complex vector is defined as:

$$\hat{Y}_m^l = \frac{\bar{Y}_m^l}{|\bar{Y}_m^l|} \qquad (2)$$

where $l = 4$. This calculation is performed in LAMMPS by using the *orientorder/atom* compute and setting the *components* value to 4. The outcome of this step is a 2×(2$l$ + 1) component array corresponding to the real and imaginary parts of normalized $\hat{Y}_m^l$. Next, the similarity between the angular distributions of a central atom $i$ and its neighboring atoms is evaluated as:

$$s_{ij} = \sum_{m=-l}^{l} \hat{Y}_m^l(i)\hat{Y}_m^{l*}(j) \qquad (3)$$

This resulting value of $s_{ij}$ is always real. Finally, the disorder parameter for each atom $i$ is computed as:

$$d_i = 1 - \frac{|s_{ij}|}{N_i} \qquad (4)$$

where a value of 0 corresponds to atoms in a perfect lattice, and a value of 1 signifies atoms in a liquid [2,3]. The excess disorder is then determined as:

$$\Gamma_{dis} \equiv (\Sigma_{i=1}^{N} d_i^{GB} - \Sigma_{i=1}^{N} d_i^{bulk})/2A \qquad (5)$$

where $\Sigma_{i=1}^{N} d_i^{GB}$ is the sum of atomic disorder over all $N$ atoms in the bicrystal sample, $\Sigma_{i=1}^{N} d_i^{bulk}$ is the corresponding sum of disorder over $N$ atoms in a simulation cell without defects, and $2A$ represents total grain boundary area.

The grain boundary free volume (GBFV) is defined as:

$$(\bar{V}_{atom}^{GB} - \bar{V}_{atom}^{bulk})/\bar{V}_{atom}^{bulk} \qquad (6)$$

where $\bar{V}_{atom}^{GB}$ is the average atomic volume of defect atoms and $\bar{V}_{atom}^{bulk}$ is the average atomic volume of atoms within the grains.

The grain boundary thickness is calculated as:

$$\sum_{i=1}^{N_{GB}} V_i/2A, \qquad (7)$$

where $\sum_{i=1}^{N_{GB}} V_i$ is the sum of atomic volumes over all $N_{GB}$ grain boundary atoms, and $2A$ again denotes the total grain boundary area.

**Supplementary Note 3**

To visually and quantifiably differentiate between the metastable and equilibrium configurations, it is essential to employ a parameter that comprehensively considers all atoms within each structure. In this work, the parameter utilized is derived from the normalized complex vector, $\hat{Y}_m^l$, as explained in Supplementary Note 2. This complex vector is composed of a 20 component array that includes both real and imaginary values.

To visualize the equilibrium structure, the real component corresponding to $\hat{Y}_{4(m=-1)}$ is calculated for each atom (Fig. S3(b)). These calculated values range from -1 to 1, with values less than 0 corresponding to unique atomic structures inherent to the equilibrium structure (Fig. S3(c)). Subsequently, a radial expansion of 5 Å is performed around these selected atoms to include neighboring atoms that contribute to the equilibrium structure (Fig. S3(d)).

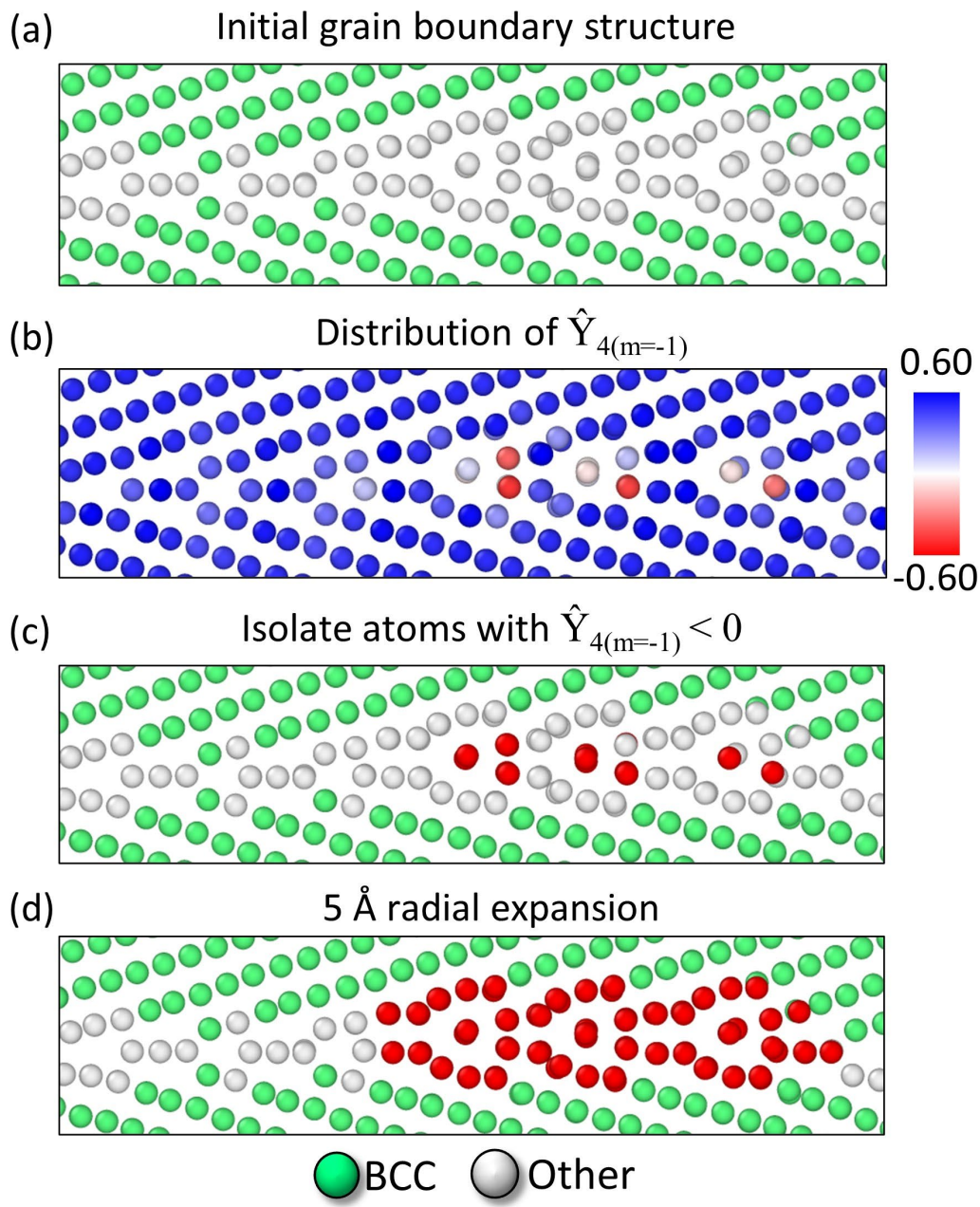

Fig. S3. Sequential steps for differentiating the metastable and equilibrium grain boundary structures. (a) The initial grain boundary structure identified using aCNA. (b) Atomic color coding corresponding to the local $\hat{Y}_{4(m=-1)}$ value. (c) Illustration of the atoms with $\hat{Y}_{4(m=-1)}$ value less than zero (highlighted in red). (d) A 5 Å radial expansion around the atoms highlighted in (c) resulting in accurate distinction between the two structures.

**Supplementary Note 4**

Fig. S4 shows the equilibrium grain boundary structure (1600 K) in pure Nb and in NbMoTaW after high-temperature annealing using the SNAP. The same structure observed in the main text is again recreated here in both pure and chemically complex environments. The chemical complexity is highlighted in the lower panel of Fig. S4 where the grain boundary and nearby crystalline regions are colored by atom type. The black circles in the two differently colored NbMoTaW images indicate the same structural unit (and same set of atoms) within the two snapshots.

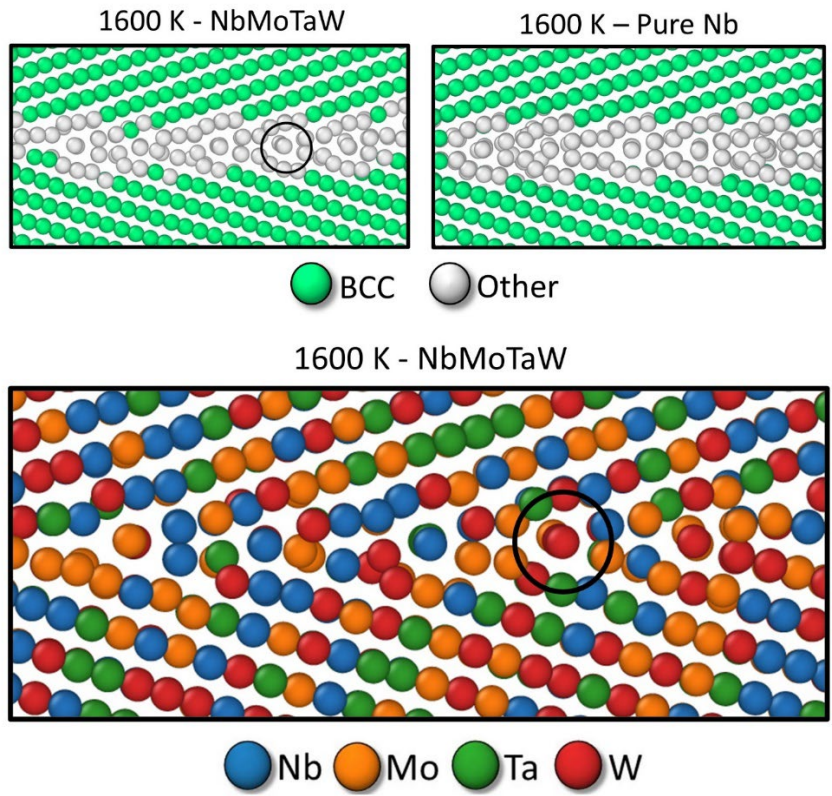

**Fig. S4. Snapshots of the grain boundary structures identified in pure Nb and NbMoTaW using SNAP. The grain boundary structure (gray atoms) at 1600 K are shown in the top panel with similar equilibrium grain boundary structures as those found using the MTP. In the lower panel, coloring by atom type highlights the**

chemical inhomogeneity at high temperatures. The black circle in both NbMoTaW samples guides the eye to identify the location of the grain boundary structural unit.

**Supplementary Note 5**

To better understand the effect of structure on chemical interactions, we assess local chemical ordering in the bulk and interfacial regions as a function of temperature by computing the first-nearest neighbor pair-wise Warren-Cowley parameter [4], defined as:

$$\alpha_n^{ij} = 1 - \frac{p_n^{j,i}}{c_j}. \tag{8}$$

The order parameter $\alpha_n^{ij}$ is a descriptor which captures element interaction probabilities between the center element species $i$ and neighboring species $j$ in the $n$th neighbor shell. In the current work, only the first nearest-neighbor shell is considered (i.e. $n=1$). In Equation (8), $p_n^{j,i}$ is the concentration of atom type $j$ around atom type $i$, and $c_j$ is the global concentration of atom type $j$. Fig. S5 displays the calculated order parameter for intra-species (e.g. Nb-Nb) and inter-species (e.g. Mo-Ta) pairs with positive and negative $\alpha_1^{ij}$ values representing unfavorable and favorable bonding probabilities in the first coordination shell.

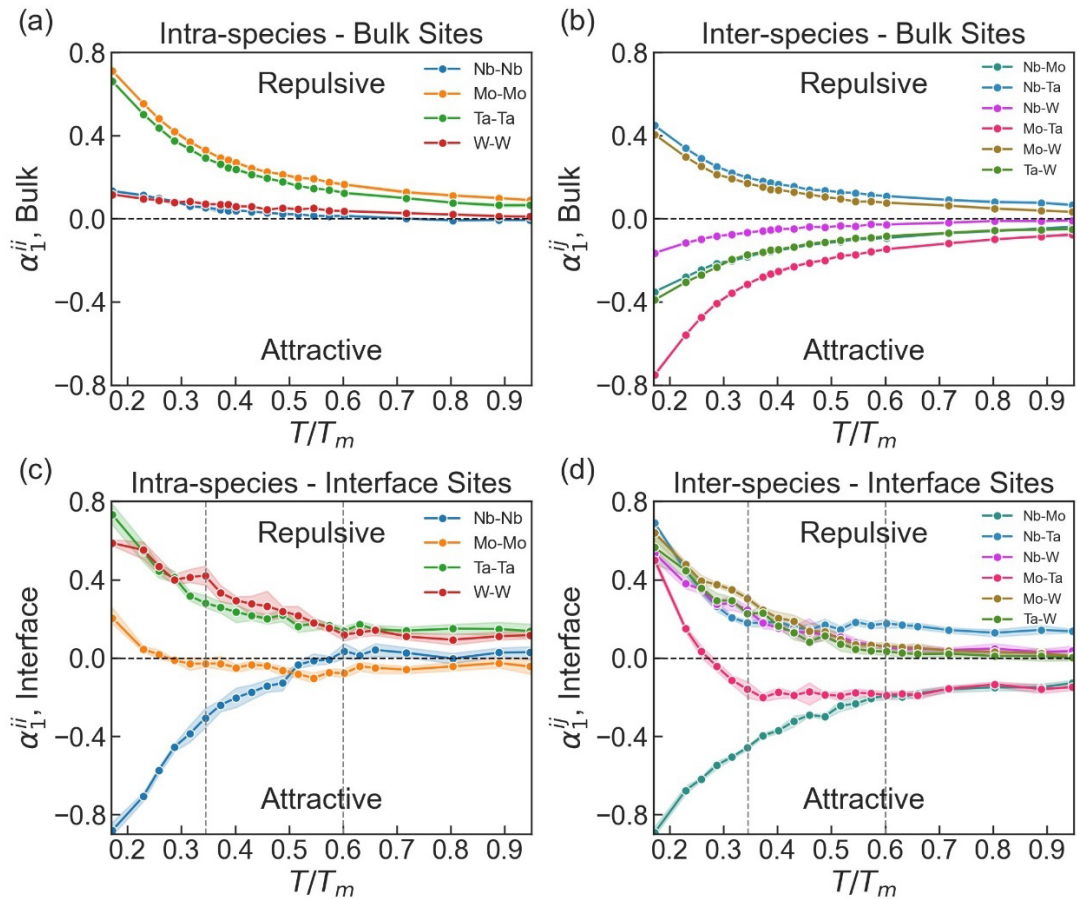

**Fig. S5.** Intra-species and inter-species chemical short-range ordering behavior calculated using the Warren-Cowley parameter for (a, b) bulk sites and (c, d) interface sites. The dashed lines in (c) and (d) indicate the transition temperature range. Note that a positive $\alpha$ value indicates repulsive interaction and a negative $\alpha$ signifies attractive interaction among the two elements.

**Supplementary Note 6**

Fig. S6 shows the mean-square displacement (MSD) values for the RSS grain boundary atoms at a low temperature of 700 K (i.e., $0.20T_m$) and each color on the plot represents a separate diffusion simulation run. For 4 out of 5 simulations, the flat MSD trend indicates that zero atomic hops occurred over the 15 ns timeframe. The slight increase in the MSD around 9 ns for the brown circles correlates with a singular (or cooperative) atomic hop(s). In Fig. S7(a) and (b), the MSD for the Segregated and RSS grain boundary atoms, respectively, is plotted as a function of time at various temperatures below the transition initiation temperature. As temperature increases, the slope of the MSD for each ordering condition also increases, indicating greater atomic mobility within the interface. In Fig. S8, the MSD values for RSS and Segregated grain boundaries without vacancies are shown at 900 K. Direct comparisons of Figs. S7 and S8 demonstrate the effect of vacancy addition on the MSD increase. Notably, much longer timescales are required to obtain a diffusivity value for the RSS state at 900 K due to the lowered probability of atomic jumps as well as the fluctuations that deviate from the expected linear behavior. For the Segregated state, no atomic motion is observed, thus underscoring the need for vacancy additions to obtain meaningful data about the relative kinetics of the two chemical configurations.

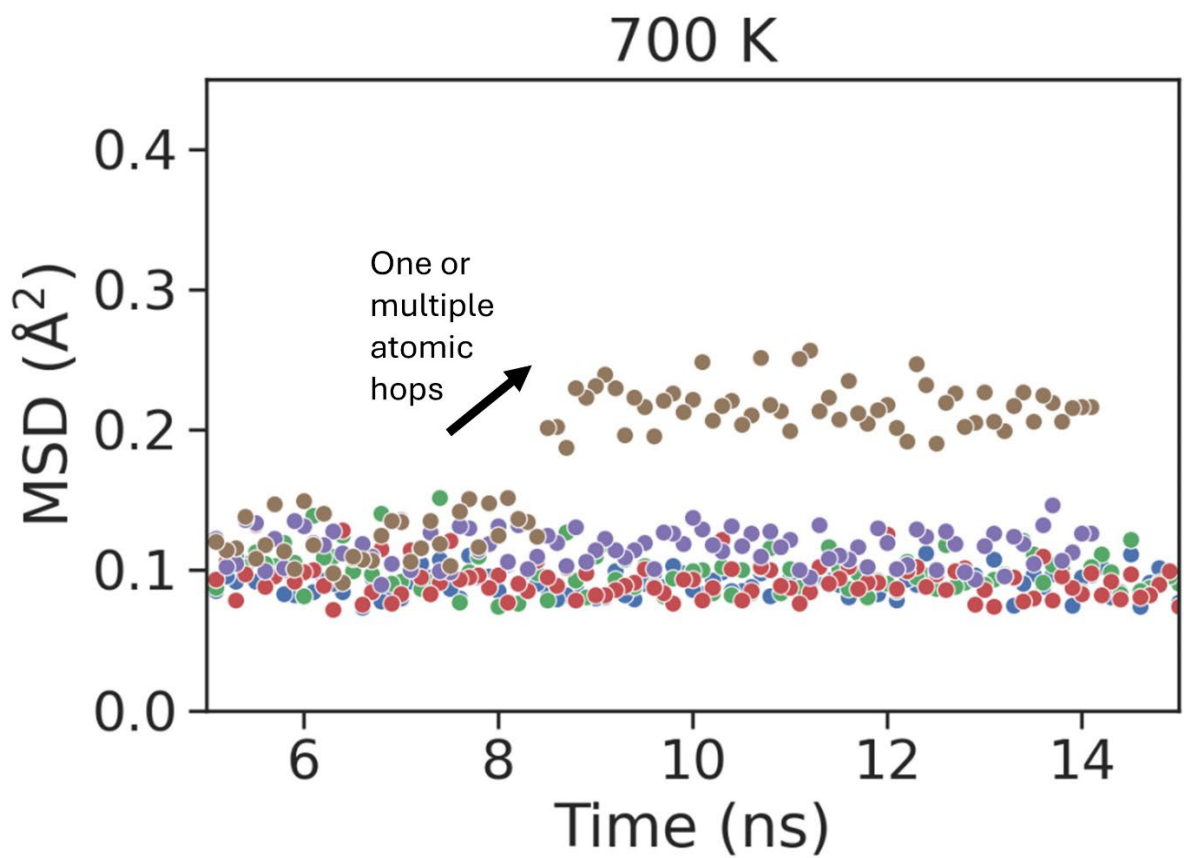

**Fig. S6.** Mean-squared displacement of grain boundary atoms as a function of time for the RSS chemical configurations performed at 700 K (0.20$T_m$). Each color represents a separate diffusion simulation. The arrow highlights one or multiple atomic hops, the sum of which typically contribute to a measured diffusivity.

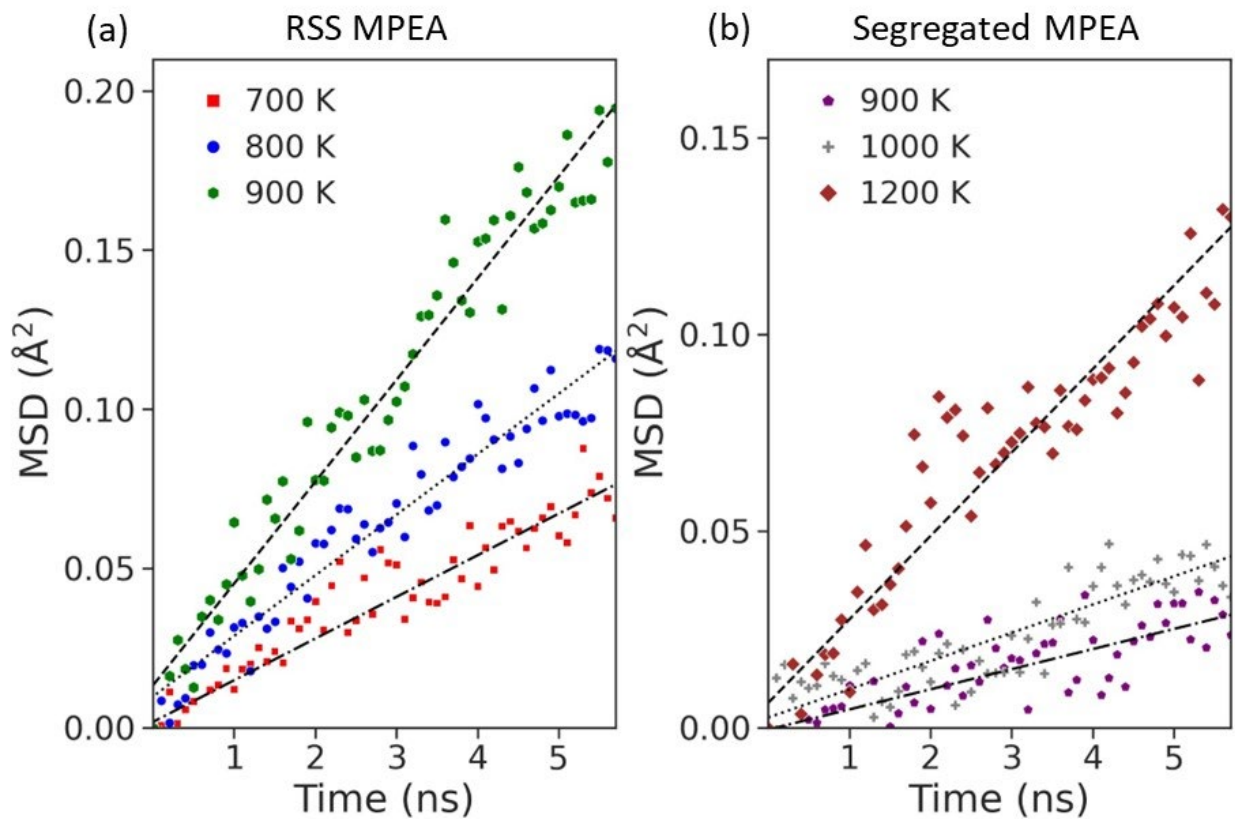

**Fig. S7. Mean-squared displacement of grain boundary atoms as a function of time for the (a) RSS and (b) Segregated chemical configurations performed at varying temperatures below their respective transition temperatures. The linear relation between MSD and time is shown for each dataset with different types of dashed lines.**

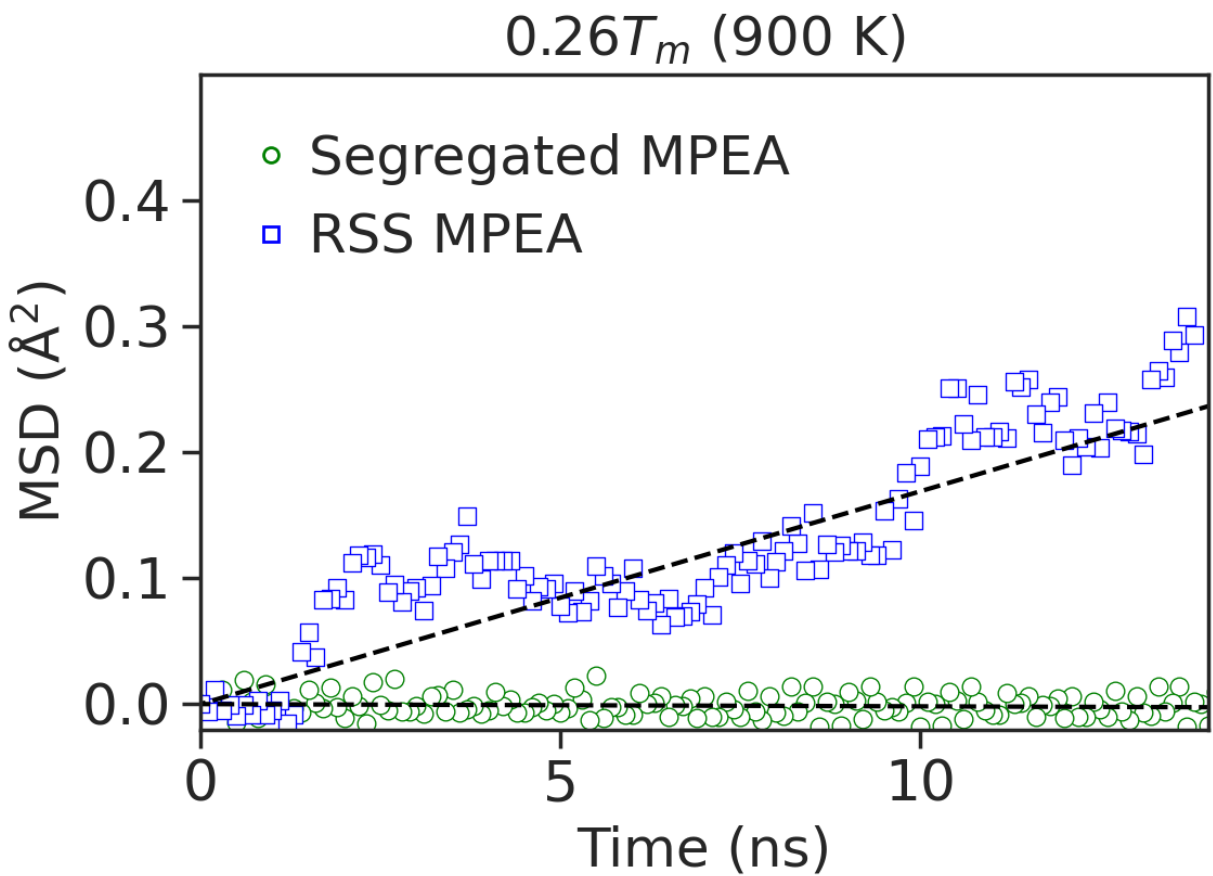

**Fig. S8. Mean-squared displacement of grain boundary atoms as a function of time for the RSS and Segregated chemical configurations performed at 900 K without the addition of vacancies. The linear relation between MSD and time is shown for each dataset with dashed lines.**